\documentclass[%
 reprint,
 amsmath,amssymb,
]{revtex4-2}

\usepackage{graphicx}
\usepackage{dcolumn}
\usepackage{bm}
\usepackage{hyperref}
\usepackage[T1]{fontenc}

\usepackage[dvipsnames]{xcolor}

\newcommand{\bn}{$ \begin{array}{l} \vspace{-0.0cm}}
\newcommand{\en}{\end{array} $}

\newcommand{\aap}{Astron. Astrophys.}
\newcommand{\apjl}{Astrophys. J. Lett.}
\newcommand{\ssr}{Space Science Reviews}
\newcommand{\nphysa}{Nucl. Phys. A}
\newcommand{\arnps}{Annu. Rev. Nucl. \& Part. Sci.}
\newcommand{\araa}{Annu. Rev. Astron. \& Astrophys.}
\newcommand{\physrep}{Phys. Rep.}
\newcommand{\mnras}{Monthly Not. Royal Astron. Soc.}

\begin{document}

\preprint{APS/123-QED}

\title{Probing Strong Field $f(R)$ Gravity and Ultra-Dense Matter with the \\ Structure and Thermal Evolution of Neutron Stars}
\author{Martin Nava-Callejas}
\email{mnava@astro.unam.mx}
\author{Dany Page}%
\email{page@astro.unam.mx}

\affiliation{%
Instituto de Astronom\'ia, Universidad Nacional Aut\'onoma de M\'exico, Ciudad de M\'exico, CDMX 04510, Mexico}%
\author{Mikhail V. Beznogov}
\email{mikhail@astro.unam.mx}
\affiliation{%
Horia Hulubei National Institute of Physics and Nuclear Engineering: Bucharest, Romania}%

\date{\today}

\begin{abstract}
Thermal evolution of neutron stars is studied in the $f(R)=R+\alpha R^{2}$ theory of gravity. 
We first describe the equations of stellar structure and evolution for a spherically symmetric spacetime plus a perfect fluid at rest.
We then present numerical results for the structure of neutron stars using four nucleonic dense matter equations of state and a series of gravity theories for
$\alpha$ ranging from zero, i.e., General Relativity, up to $\alpha \approx 10^{16}$ cm$^2$.
We emphasize properties of these neutron star models that are of relevance for their thermal evolution as the threshold masses for enhanced neutrino emission
by the direct Urca process, the proper volume of the stellar cores where this neutrino emission is allowed, the crust thickness, 
and the surface gravitational acceleration that directly impact the observable effective temperature.
Finally, we numerically solve the equations of thermal evolution and explicitly analyze the effects of altering gravity.
We find that uncertainties in the dense matter microphysics, as the core chemical composition and superfluidity/superconductivity properties, as well as the 
astrophysical uncertainties on the chemical composition of the surface layers, have a much stronger impact than possible modifications of gravity 
within the studied family of $f(R)$ theories.
We conclude that within this family of gravity theories, conclusions from previous studies of neutron star thermal evolution are not significantly
altered by alteration of gravity.
Conversely, this implies that neutron star cooling modeling may not be a useful tool to constrain deviations of gravity from Einstein theory
unless these are much more radical than in the $f(R)=R+\alpha R^{2}$ framework.
\end{abstract}

\maketitle

\section{Introduction}
\label{Sec:Intro}

Neutron stars contain one of the densest  forms of matter and strongest gravitational fields in the observable Universe.
Possibly higher densities have been reached in relativistic heavy ion collisions but only in a short-lived unstable form and with extremely high energy densities \cite{Busza:2018aa}, while the densest form of matter in the interior of black holes is not observable.
Neutron stars can also be copious neutrino emitters \cite{Burrows:1986ui,Yakovlev:2001wb}
and some exhibit the strongest known magnetic fields \cite{Kaspi:2017ts}.
While the strongest gravitational potentials are found near the horizon of black holes, 
higher curvatures are present within neutron stars due to their smaller radii \cite{Psaltis:2008vz}.
As such, they are prime candidates to study the interplay between the four fundamental forces of Nature.

Immediately after A. Einstein presented his final formulation of the field equations of General Relativity (GR) in 1915 \cite{Einstein:1915aa,Einstein:1915ab},
D. Hilbert showed that these can be very elegantly deduced from a minimal action principle assuming a Lagrangian density  given by the Ricci scalar $R$ \cite{Hilbert:1915un}.
Extensions of GR by replacing $R$ with a more general function $f(R)$ can be traced back to H. Weyl who already in 1918  proposed an $R^2$ version of gravity theory \cite{Weyl:1918tp}.
Among the zoo of possible extensions of gravity theory beyond GR \cite{Capozziello:2011uy}, $f(R)$ theories have remained popular as they provide a conceptually
simple scheme with minimal departure from the very successful Einstein theory \cite{Sotiriou:2010wq,De-Felice:2010vl}.
However, going beyond $f(R)=R$ already has the enormous cost that the field equations are not anymore of second order.
When dealing with the structure of a spherical star, the result is that the Tolman-Oppenheimer-Volkoff equations of hydrostatic equilibrium, which are a simple set
of 3 first order ordinary differential equations in GR (see, e.g., \cite{Shapiro:1983wz}) that are straightforward to integrate, are now replaced by much more complicated equations.
To date, many models of neutron star in $f(R)$ gravity have finally been published, some representative results being presented, e.g., in \cite{Yazadjiev_2014} for non rotating stars,
including fast rotating ones \cite{Yazadjiev:2015wu}.
Models of neutron star evolution in $f(R)$ gravity have been recently presented \cite{Dohi:2021uq}.

On the other side, the first models of neutron star looked for analytical solutions \cite{Tolman:1939tj}
or considered matter constituted by free neutrons, i.e., without any interaction \cite{Oppenheimer:1939vk},
while modern theories of dense matter take into account the best possible models of nuclear interaction calibrated to laboratory data and, at increasing densities,
the likely appearances of hyperons, meson condensates and/or deconfined quark matter are considered (see, e.g., \cite{Page:2006tr} and \cite{Burgio:2021vj}, among many others, 
for reviews). 
The resulting equation of state (EOS), i.e., the function $P=P(\rho)$ where $P$ is the pressure and $\rho$ the mass density, can be probed by 
calculating the structure of neutron stars and comparing the resulting mass, $M$, versus radius, $r_{\ast}$, relationships, $M=M(r_{\ast})$, with astrophysical data \cite{Lattimer:2007vu,Lattimer:2012vd,Ozel:2016tm}.

More intimate properties of a system, beyond its ground state properties, are revealed by its excitation levels that manifest themselves at finite temperatures.
Studies of the thermal evolution of neutron stars open a window on such finite temperature properties as
transport coefficients, neutrino emission, and specific heat \cite{Yakovlev:2004iq,Page:2006vt,Potekhin:2015tm}.
For example, the occurrence of fast neutrino emission by processes as the direct Urca one, in which neutrinos are emitted when baryons undergo circular beta and inverse beta
transformations, provides us with information about the details of the chemical composition of the matter as the proton abundance \cite{Boguta:1981te,Lattimer:1991wi},
or the presence of constituents beyond nucleons as meson condensates \cite{Maxwell:1977ug,Brown:1988aa,Tatsumi:1988wv},
hyperons \citep{Prakash:1992aa}, or deconfined quark matter \citep{Iwamoto:1980aa}.
Moreover, the single particle excitation spectrum of a baryonic constituent can be strongly modified by pairing, as, e.g., in neutron superfluidity or proton superconductivity,
with dramatic consequences on the thermal evolution of the star (see, e.g., \cite{Page:2014aa}).

Our focus in the present work will precisely be the interplay between the structure and evolution of neutron stars and strong field gravity.
We see it as a two way issue: how can we use what we know and understand, mitigated by what we do not know and/or do not understand, about neutron stars
to constrain strong field gravity and, reciprocally, how do uncertainties in our description of strong field gravity limit our possibilities to constrain the properties of dense matter by studying neutron stars.
We will restrict ourselves to modified (compared to GR) gravity theories with $f(R)=R+\alpha R^{2}$.
However, we consider that the function $f(R)$ does not necessarily have a single analytical expression representing it in the whole possible range of $R$ values.
We thus take the position that the simple analytical expression $f(R)=R+\alpha R^{2}$ represents $f(R)$ in the extreme environment of a compact star 
while a different expression could describe it in the context of the Solar system and on Earth, or at galactic and cosmological scales.
As such, constraints from cosmology, the Solar system or the Eöt-Wash experiment \cite{Berry:2011vg}, all in the context of $R \rightarrow 0$,
are not expected to constrain the $\alpha$ of our $f(R)$.

As is well known, it is possible in principle to rewrite the $f(R)$ Lagrangian density as a scalar field theory one by defining an auxiliar field $\Phi_{\text{aux}} = df/dR$ and an associated potential $V(\Phi_{\text{aux}})$. At that point, the usage of conformal transformations becomes frequent as it allows to (a) recast the Lagrangian density as $\mathcal{L}\sim R^{\ast} + \mathcal{L}(\Phi^{\ast}_{\text{aux}})$, i.e., GR with an appropriate scalar curvature $R^{\ast}$ and conformally transformed scalar-field $\Phi^{\ast}_{\text{aux}}$, (b) deduce equations of motion which are, in some sense, easier to handle numerically \cite{Capozziello:2011uy, Yazadjiev_2014, Yazadjiev:2015wu, Dohi:2021uq}. The turn-back of this approach is the necessity of clarifying what we must understand as \textit{physical equivalence} of the field equations and/or the observable quantities. Thus, we opt for analyzing the original Lagrangian density, its equations of motion and physical predictions without re-defining fields or using conformal transformations.

The structure of this paper is as follows: 
in Section ~\ref{Sec:f(R)} the field equations of the $\alpha R^{2}$ theory are reviewed, considering a perfect fluid and a static and spherically symmetric spacetime. 
In Section ~\ref{sect:cooleqs}, the essential aspects of the energy balance and heat transport equations are discussed. 
The numerical implementation of the equations and the results are given in Sections ~\ref{Sec:Structure} and ~\ref{Sec:Cooling}.
We discuss our results in Section ~\ref{sec:discussion} and briefly conclude in Section ~\ref{sec:conclusions}.

\section{The $f(R)$ Gravity Theory and the Stellar Structure Equations}
\label{Sec:f(R)}

To avoid ambiguities we first summarize our geometric conventions.
We adopt the metric signature $(-,+,+,+)$ and take the Riemann tensor as
\begin{equation}
R^{\alpha}_{\ \beta\mu\nu} = \partial_{\mu}\Gamma^{\alpha}_{\ \beta\nu}-\partial_{\nu}\Gamma^{\alpha}_{\ \beta\mu}+\Gamma^{\alpha}_{\ \sigma\mu}\Gamma^{\sigma}_{\ \beta\nu}-\Gamma^{\alpha}_{\ \sigma\nu}\Gamma^{\sigma}_{\ \beta\mu}
\end{equation}
where $\partial_{\mu}$ denotes partial derivation with respect to the coordinate $x^{\mu}$ and $\Gamma^{\alpha}_{\mu\nu}$ is the Levi-Civita connection
\begin{equation}
\Gamma^{\alpha}_{\ \mu\nu} = \frac{1}{2}g^{\alpha\varepsilon}\left(\partial_{\mu}g_{\nu\varepsilon}+\partial_{\nu}g_{\mu\varepsilon}-\partial_{\varepsilon}g_{\mu\nu}\right)\ ,
\end{equation}
which serves to define the covariant derivative $\nabla_{\mu}$ for this work as, e.g., for a rank two tensor $T^\alpha_\beta$,
\begin{equation}
\nabla_{\mu} T^{\alpha}_{\ \beta} = \partial_\mu T^{\alpha}_{\ \beta}+\Gamma^{\alpha}_{\ \sigma\mu}T^{\sigma}_{\ \beta} - \Gamma^{\sigma}_{\ \mu\beta}T^{\alpha}_{\ \sigma}\ .
\end{equation}
The contraction $R^{\alpha}_{\ \beta\alpha\nu} = R_{\beta\nu}$ is the Ricci tensor, its trace $R=g^{\alpha\beta}R_{\alpha\beta}$ is the scalar curvature and the combination 
\begin{equation}
G_{\mu\nu} = R_{\mu\nu} - \frac{1}{2}g_{\mu\nu}R
\end{equation}
is the Einstein tensor. Regarding the D'Alembertian operator applied to a scalar field $\phi$, it is defined by
\begin{equation}
\Box\phi = \nabla^{\mu}\nabla_{\mu}\phi = \frac{1}{\sqrt{-g}}\partial_{\nu}\left[\sqrt{-g}g^{\mu\nu}\partial_{\mu}\phi\right]\ ,
\end{equation}
with $g$ the determinant of the metric tensor. Finally, for single-variable functions we employ either $df(x)/dx$ or $f'$ to denote derivation with respect to the argument $x$.

Regarded as an effective field theory for gravity, the motivation for this theory consists in the replacement of the standard Hilbert-Einstein Lagrangian density 
$\mathcal{L}\propto R$ with a function of the scalar curvature $f(R)$, which in order to be ghost-free must satisfy \cite{Yazadjiev_2014}:
\begin{equation}
\frac{df}{dR}\geq 0\ ,\ \frac{d^{2}f}{dR^{2}}\geq 0\ .
\end{equation}
Being an extension of GR, gravity is exclusively described by the metric tensor, whose dynamics is enclosed in the action functional which, for the particular choice of $f(R)=R+\alpha R^{2}$, is
\begin{equation}
S = \int \sqrt{-g} \, d^{4}x\frac{1}{c}\left[\frac{c^{4}}{16\pi G}(R+\alpha R^{2}) + \mathcal{L}_{M}\right]
\end{equation}
where $c$ is the speed of light in vacuum, $G$ is Newton's constant, $\alpha$ a non-negative real number with dimensions of length squared, and $\mathcal{L}_{M}$ is the matter lagrangian. By performing a null variation with respect to $g^{\mu\nu}$, i.e $\delta S=0$, we arrive to the field equations \cite{Jaime:2011tc}, \cite{Feola:2020tw}:
\begin{eqnarray}
(1+2\alpha R) G_{\mu\nu} + \frac{\alpha R^{2}}{2} g_{\mu\nu}= 
\nonumber \\
2 \alpha \left[ \nabla_{\mu}\nabla_{\nu} - g_{\mu\nu} \Box \right] R + \frac{8\pi G}{c^{4}} T_{\mu\nu}
\end{eqnarray}
where $T_{\mu\nu}$ is the energy-momentum tensor \cite{2021IJGMM..1850059M}
\begin{equation}
T_{\mu\nu} = -\frac{2}{\sqrt{-g}}\frac{\delta(\sqrt{-g}\mathcal{L}_{M})}{\delta g^{\mu\nu}} 
\end{equation}
which, for this theory of gravity, still satisfies the conservation law
\begin{equation}
\nabla_{\mu}T^{\mu\nu} = 0\ .
\label{eq:emconserv}
\end{equation}
In contrast with GR, the trace $T=g^{\mu\nu}T_{\mu\nu}$ and the scalar curvature $R$ are related by a differential expression, 
\begin{equation}
\left(6\alpha\Box-1\right) R = \frac{8\pi G}{c^{4}}T \ ,
\label{eq:fieldfr}
\end{equation}
which constitutes an additional relation for the variables of the model.

To study non-rotating neutron stars, we model the metric as static and spherically symmetric (SSS):
\begin{equation}
ds^{2} = -e^{2\Phi(r)}c^{2}dt^{2} + e^{2\Lambda(r)}dr^{2}+r^{2}d\Omega^{2}\ ,
\label{Eq:metric}
\end{equation}
where $d\Omega^{2} = d\theta^{2}+\sin^{2}\theta d\phi^{2}$. 
The coordinates in this metric have exactly the same interpretation as in GR.
The 3D metric on a hypersurface (sphere) $r=$ constant is identical to the one in Euclidean space, in particular the area of this sphere
is $S(r) = 4\pi r^2$ and $r$ can be called the "areal radius".
The proper radial length, $dl$, measured by a local observer at rest on the shell is related to $dr$ by $dl = \exp[\Lambda(r)] \, dr$ 
while his proper time interval $d\tau$ is given by $d\tau = \exp[\Phi(r)] \, dt$.
We also impose the condition that the metric tends to \textit{Schwarzschild's}
\begin{eqnarray}
\Lambda(r)+\Phi(r) = 0\ ,\nonumber\\
e^{2\Phi(r)} = \left[1-\frac{2GM}{c^{2}r}\right]^{-1}\ ,
\end{eqnarray}
as $r\to\infty$.
Hence, $t$ can  be interpreted as the proper time of an observer at rest at infinity.

For neutron stars, we are working under the assumption that they are adequately described by a perfect fluid defined by its pressure $P$ and 
energy density $\varepsilon$ providing a simple energy momentum tensor
\begin{equation}
T_{\mu\nu} = (\varepsilon+P)u_{\mu}u_{\nu} + Pg_{\mu\nu}\ ,
\end{equation}
with $u^{\mu}$ the fluid rest-frame unitary 4-velocity, $u^{\mu}u_{\mu}=-1$.
By inserting this tensor and the SSS metric in the field equations, we obtain the following set of second-order, non-linear differential equations \cite{Feola:2020tw}
\begin{equation}
\frac{d\Phi}{dr} = \frac{1}{A_{1}}\left\{\frac{4\pi GPre^{2\Lambda}}{c^{4}}-\frac{[A_{2}e^{2\Lambda}+2A_{3}]}{4r}\right\}
\label{eq:structure1}
\end{equation}
\begin{eqnarray}
\!\!\!\!\!\!\!\!\!\!\!\!\frac{d\Lambda}{dr} = \frac{1}{A_{1}}\left\{\frac{4\pi G\varepsilon re^{2\Lambda}}{c^{4}}+\frac{[A_{2}e^{2\Lambda}+2A_{3}]}{4r}+\alpha r\frac{d^{2}R}{dr^{2}}\right\}
\label{eq:structure2}
\\
\!\!\!\!\!\!\!\!\!\!\!\!\frac{d^{2}R}{dr^{2}} = \frac{1}{A_{6}} \!\! \left\{\!\! \frac{A_{1}A_{4}e^{2\Lambda}}{6\alpha}\!-\!\!\frac{1}{r}\!\!\left[A_{5}-\frac{A_{2}e^{2\Lambda}}{2}\right]\!\!\frac{dR}{dr}\right\}
\label{eq:structure3}
\end{eqnarray}
where, for simplicity of notation, the following functions have been defined:
\begin{equation}
A_{1}=1+2\alpha R+\alpha r\frac{dR}{dr}\ ,\ A_{2}=\alpha r^{2}R^{2}-4\alpha R-2\ ,
\end{equation}
\begin{equation}
A_{3}=1+2\alpha R+4\alpha r\frac{dR}{dr}\ ,\ A_{4}=\frac{8\pi G}{c^{4}}(3P-\varepsilon)+R\ ,
\end{equation}
\begin{equation}
A_{5}=\frac{4\pi G}{c^{4}}r^{2}e^{2\Lambda}(P-\varepsilon)+1+2\alpha R-2\alpha r\frac{dR}{dr}\ ,
\end{equation}
\begin{equation}
A_{6}=1+2\alpha R\ .
\end{equation}
From the $r$-component of the energy-momentum tensor conservation law, Eq.~\ref{eq:emconserv}, we deduce the differential equation for the pressure,
\begin{equation}
\frac{dP}{dr} = -(P+\varepsilon)\frac{d\Phi}{dr}\ .
\label{eq:structure4}
\end{equation}

We thus have five structural functions, $\Phi(r)$, $\Lambda(r)$, $R(r)$, $P(r)$, and $\varepsilon(r)$, and four equations, 
~\ref{eq:structure1}, ~\ref{eq:structure2}, ~\ref{eq:structure3}, and  ~\ref{eq:structure4},
which need to be completed by an Equation of State (EoS) giving a relationship of the form $P=P(\varepsilon)$ \cite{Page:2006tr}.
Instead of $\varepsilon$ it is also customary to employ the mass density $\rho$ such that $\varepsilon = \rho c^2$.
Microscopic calculations of the EoS and astrophysical processes in general are commonly
phrased in terms of baryons through their correspondent density $n$, such that $P=P(n)$ and $\varepsilon=\varepsilon(n)$. 
For our study the temperature dependence of the EoS can be neglected and we do not need further equations to determine the structure
of the star, except for the outer layers of the envelope as discussed below in \S~\ref{Sec:Cooling_Numeric}.
 
For compact objects, the exterior is defined as the region where both pressure and energy density are identically zero. 
From the differential equations, it is clear that these conditions do not imply a vanishing scalar curvature outside the objects, 
but it is nevertheless possible to demand it as an asymptotic limit, recovering Schwarzschild's metric in the process. 
Indeed, from the differential equations for $\Lambda$ and $\Phi$ we have
\begin{equation}
\frac{d\Phi}{dr}+\frac{d\Lambda}{dr} = \frac{\alpha r}{A_{1}}\frac{d^{2}R}{dr^{2}}\ .
\label{eq:lambplusphi}
\end{equation}
For the left hand side to vanish, $d^{2}R/dr^{2} = 0$, which implies that $R=0$ is a possible asymptotic solution. 
Aside from this fact, let us stress that the treatment of $R$ as an independent variable has the advantage of keeping a second order system of equations instead of a fourth-order one: 
applying the geometric definition $R=g^{\mu\nu}R_{\mu\nu}$ to the SSS metric leads to $R\propto\Phi''$, hence $R''\propto\Phi^{(4)}$.
Notice that in GR the scalar curvature is given by
\begin{equation}
R^{(\mathrm{GR})} = \frac{8\pi G}{c^4} (\varepsilon-3P)
\label{eq:R_GR}
\end{equation}
and naturally vanishes outside the star.
In the center of a neutron star values of $|R^{(\mathrm{GR})}|$ of the order of $10^{-12}$ cm$^{-2}$ are found.

We will describe the boundary conditions needed to solve these structure equations in details in \S~\ref{Sec:Structure_Numeric}.

\subsection{Terminology and notations}
\label{sect:notations}

We will simply refer to this theory of gravity with $f(R) = R + \alpha R^{2}$ as the ``$\alpha R^2$ theory'' or as ``$\alpha R^2$ gravity''.
For the parameter $\alpha$, we adopt the following notation:
\begin{equation}
\alpha_{X} = \left(\frac{GM_{\odot}}{c^{2}}\right)^{2}\times 10^{X-10} = \text{2.18}\times 10^{X}\ \text{cm}^{2}\ ,
\end{equation}
considering for this work the cases $X=9,10,11,12,14$, and $16$.

\section{The Stellar Evolution Equations}
\label{sect:cooleqs}

\begin{table*}[t]
\begin{center}
\caption{Examples of neutrino emitting processes in neutron star cores$^1\! $.}
\label{Tab:nu}
\begin{tabular}{llcc} 
\noalign{\smallskip}\hline
    Name       &         Process        &       Emissivity$^2$ $q_\nu$             &        \\  
               &                        &   (erg cm$^{-3}$ s$^{-1}$)       &        \\
\noalign{\smallskip}\hline
\parbox[c]{2.6cm}{Modified Urca\\ (neutron branch)} &
\rule[-0.30cm]{0.01cm}{0.70cm}
\bn n+n \rightarrow n+p+e^-+\bar\nu_e \\ n+p+e^- \rightarrow n+n+\nu_e \en  & 
$\sim 2\!\!\times\!\! 10^{21} \: {\cal R} \: T_9^8$ & Slow \\
\parbox[c]{2.6cm}{Modified Urca\\ (proton branch)}  &
\rule[-0.30cm]{0.01cm}{0.70cm}
\bn p+n \rightarrow p+p+e^-+\bar\nu_e \\ p+p+e^- \rightarrow p+n+\nu_e \en  & 
$\sim 10^{21} \: {\cal R} \: T_9^8$ & Slow \\
Bremsstrahlung          &
\bn n+n \rightarrow n+n+\nu \bar\nu \\ 
    n+p \rightarrow n+p+\nu \bar\nu \vspace{-0.0cm} \\
    p+p \rightarrow p+p+\nu \bar\nu \en                                     & 
$\sim 10^{19} \: {\cal R} \: T_9^8$  & Slow \\
\parbox[c]{2.6cm}{Cooper pair \\ breaking and formation$^3$}          &
\bn    n+n \rightarrow [nn] +\nu \bar\nu \\ 
       p+p \rightarrow [pp] +\nu \bar\nu \en  & 
\bn    \sim 5\!\!\times\!\! 10^{21} \: {\cal R} \: T_9^7 \\ 
       \sim 5\!\!\times\!\! 10^{19} \: {\cal R} \: T_9^7 \en & Medium  \\
Direct Urca        & 
\rule[-0.30cm]{0.01cm}{0.70cm}
\bn n \rightarrow p+e^-+\bar\nu_e \\ p+e^- \rightarrow n+\nu_e \en              & 
$\sim 10^{27} \: {\cal R} \: T_9^6$ & Fast \\
\noalign{\smallskip}\hline
\end{tabular}
\end{center}
{\footnotesize $^1\! $ Table from \cite{Page:2006vt}.}\\
{\footnotesize $^2$ $T_9$ denotes $T$ in units of $10^9$ K and
for each process the ``control coefficient'' ${\cal R}={\cal R}(T/T_\mathrm{c})$ 
is introduced to take into account the extra temperature dependence due to pairing.
Each process has its own corresponding $\cal R$.} \\
{\footnotesize $^3 $ $[nn]$ and $[pp]$ denote neutron and proton Cooper pairs..}
\end{table*}

The equations describing the thermal evolution of a spherically symmetric star in the metric (\ref{Eq:metric}) depend only on the form of this SSS metric
and not on the particular theory of gravity, within certain conditions (see last paragraphs of this section).
They are derived in all details, within the GR framework, in Ref.~\cite{Thorne:1965ww,Thorne:1967ww} and summarized in \cite{Thorne:1977tf}
and we here sketch their derivation to convince the reader that the very same equations are still valid as long as the SSS metric of Eq.~\ref{Eq:metric} is employed.

Let $T(r)$ be the temperature on the shell of areal radius $r$ and $L(r)$ the diffusive luminosity at the same position.
Both $T$ and $L$ refer to the physical quantities measured by a local observer at rest on the said shell.
In flat space-time $L$ can be assumed to be related to the gradient of $T$ through the Fourier law
\begin{equation}
L = - 4\pi r^2 K \frac{dT}{dr}
\label{Eq:dTdr1}
\end{equation}
where $K$ is the thermal conductivity in the shell (as it would be measured by a local observer at rest).
Still in Minkowsky space-time, energy conservation in this shell is expressed by
\begin{equation}
\frac{dL}{dr} = 4\pi r^2 \left[ q_h - q_\nu - n \frac{d(\varepsilon/n)}{dt} + \frac{P}{n} \frac{dn}{dt} \right]
\label{Eq:dLdr1}
\end{equation}
where $q_\nu$ is the neutrino emissivity, i.e., the energy loss per unit time and unit volume due to neutrino emission and similarly $q_h$
represent heat injection by such processes as nuclear reactions, magnetic field decay, friction due to differential rotation, etc ...
When transcribing these two equations to curved space-time, $dr$ and $dt$ in Eqs.~(\ref{Eq:dTdr1}) and (\ref{Eq:dLdr1}) are proper quantities 
and the derivative must thus we replaced as
\begin{equation}
\frac{d}{dt} \longrightarrow e^{-\Phi(r)} \frac{d}{dt} 
\;\;\;\; \text{and} \;\;\;\; 
\frac{d}{dr} \longrightarrow e^{-\Lambda(r)} \frac{d}{dr} 
\label{Eq:NewtoGR}
\end{equation}
while the factors $4\pi r^2$ remain without change since $r$ is an areal coordinate.
A crucial point is that in the presence of gravity conservation of energy must include the gravitational energy and hence, 
in the space-time with metric (\ref{Eq:metric}) the conserved energy is the {\it red-shifted} energy.
If $E$ is some energy measured at $r$ by some local observer at rest, the corresponding red-shifted energy is $\widetilde{E} \equiv \exp[\Phi(r)] E$.
A radial derivative $dE/dr$ involves comparing the energies $E(r+\delta r)$ and $E(r)$: $E(r+\delta r)$ must be moved from $r+\delta r$ to $r$ and become
$\hat{E} = \exp[\Phi(r+\delta r)-\Phi(r)] E(r+\delta r)$ and the radial derivative becomes
\begin{equation}
\frac{dE}{dr} \longrightarrow \lim_{\delta r \rightarrow 0} \frac{\hat{E}-E(r)}{\delta r} = e^{-\Lambda} \frac{d \left(e^{\Phi} E\right)}{dr} \; .
\label{Eq:dEdr}
\end{equation}
This same rule applies to $T$ in  Eq.~(\ref{Eq:dTdr1}) since it represents the energy of the diffusing particles that generate $L$.
For the derivative of $L$ in Eqs.~(\ref{Eq:dLdr1}), however, two red-shift factors are necessary because $L$ is energy per unit time and time
is itself blue-shifted.
With these consideration it is easy to verify that the stellar evolution equations in the curved space-time described by the SSS metric (\ref{Eq:metric}) are %
\begin{equation}
\frac{d(e^{\Phi}T)}{dr} = -\frac{1}{K} \frac{e^{\Phi+\Lambda}L}{4\pi r^2}
\label{Eq:dTdr_GR}
\end{equation}
and
\begin{eqnarray}
\frac{d(e^{2\Phi}L)}{dr} = 4\pi r^2\, e^{\Lambda}\times
\nonumber \\
     \left[ e^{2\Phi}q_h - e^{2\Phi}q_\nu - e^{\Phi}n \frac{d(\varepsilon/n)}{dt} + \frac{e^{\Phi}P}{n} \frac{dn}{dt} \right]
\label{Eq:Ebalance_GR}
\end{eqnarray}
which have exactly the same form as in the well known GR formulation.
We will describe the boundary conditions needed to solve these evolution equations in details in \S~\ref{Sec:Cooling_Numeric}.

Apparently, the energy balance and heat transport equations are insensible to the gravity theory chosen. 
This holds only if certain conditions are met: first, an equivalence principle must exist in the gravity theory under consideration. 
Second, if additional fields (either scalar, vector or tensor ones) are included in the Lagrangian density (which is not the case for the present work), 
it must be verified that no thermodynamical consequences arise from their introduction. 
Finally, if conformal transformations are employed (either to simplify the task of solving the field equations or to invoke the analogy with a scalar field), 
both structure and thermal differential equations must be solved in the same frame. 
In the present work, all the equations and analysis are carried out without invoking conformal transformations.

\subsection{Physical Ingredients}
\label{Sec:Cooling_Ingredients}

In the interior where matter is degenerate we can simplify the term $n\, d(\varepsilon/n)dt$ in Equation~\ref{Eq:Ebalance_GR} as
\begin{equation}
n \frac{d(\varepsilon/n)}{dt} = c_v \frac{dT}{dt} 
\label{Eq:cv1}
\end{equation}
with
\begin{equation}
c_v = n  \frac{d(\varepsilon/n)}{dT}
\label{Eq:cv}
\end{equation}
being the specific heat, per unit volume, at constant volume.
Notice that in degenerate matter there is no need to distinguish $c_v$ from the specific heat at constant pressure $c_P$ \cite{Baym-G.:2004ts}.
The microphysics ingredients, specific heat $c_v$, thermal conductivity $K$ and neutrino emissivity $q_\nu$, we employ are standard and described
in \cite{Page_2004,Page:2009um}.

For completeness we briefly present in Table~\ref{Tab:nu} the dominant neutrino emission processes occurring in the core that we consider
with schematic expressions for their emissivities and mark them as either ``slow'', ``medium'', or ``fast'' to briefly qualify their efficiency. 
The direct Urca (DUrca) process, being a highly efficient mechanism of neutrino emission,
plays a fundamental role in the cooling of NSs \cite{Boguta:1981te,Lattimer:1991wi,Page:1992tz}.
This process is only possible when the proton fraction is high enough, a criterium that can simply be stated as
\begin{equation}
n^{1/3}_{\text{p}}+n^{1/3}_{\text{e}}\geq n^{1/3}_{\text{n}}
\label{eq:DU}
\end{equation}
where $n_\text{n}$, $n_\text{p}$, and $n_\text{e}$ are the (number) densities of neutrons, protons, and electrons, respectively.
This criterium is very restrictive and is only satisfied at high densities with a threshold density $\rho_\text{DU}$ which is dependent on the EoS.

\begin{figure}[hb]
	\includegraphics[width=0.8\columnwidth]{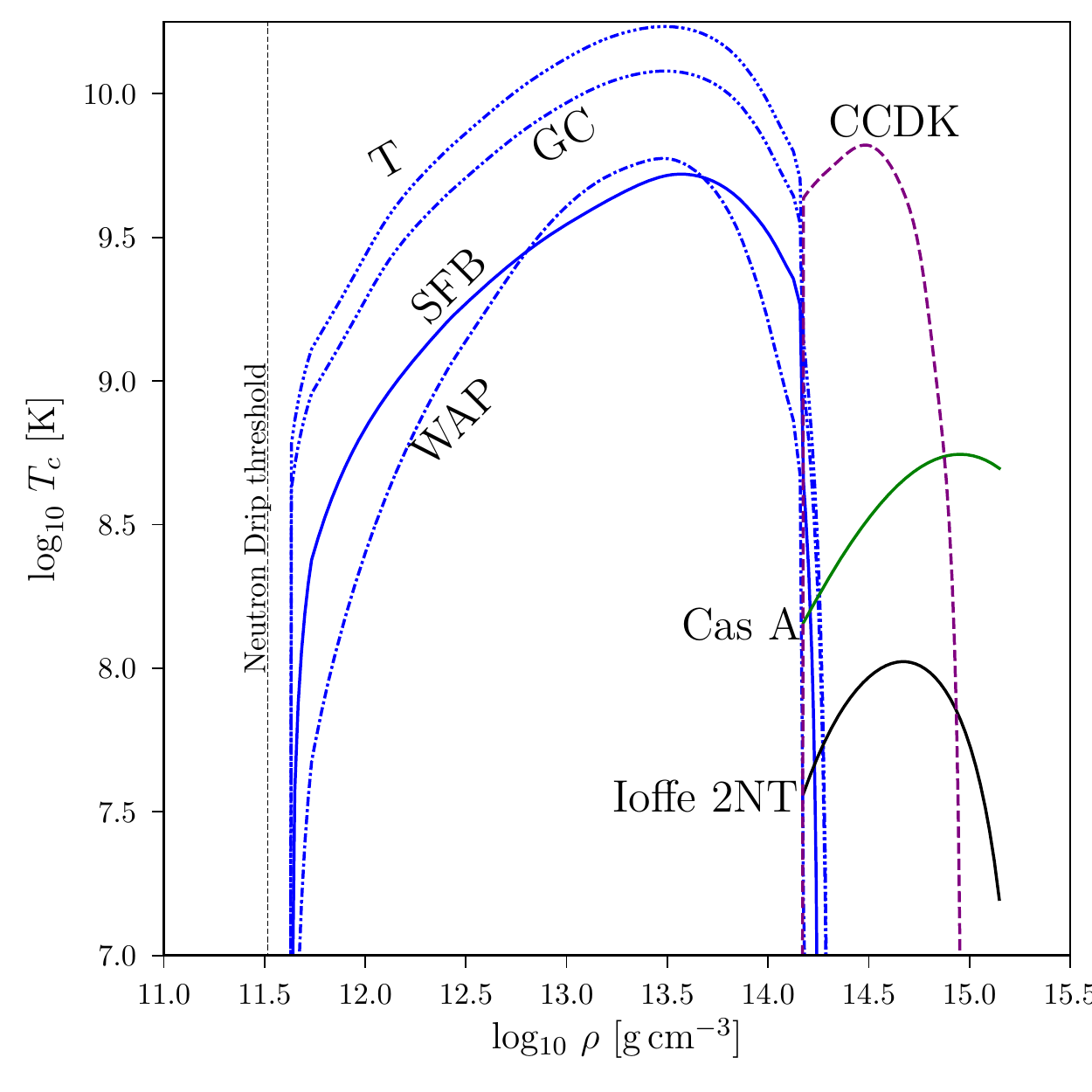} 
	\caption{$T_{c}$ vs $\rho$ for the nucleon pairing gaps considered in this work: 
	SFB \cite{Schwenk:2003vg}, WAP \cite{Wambach:1993uz}, GC \cite{Gezerlis:2008wh} and T \cite{Takatsuka:1972wy} for $^{1}\! S_{0}$ neutrons; 
	Min~a \cite{Page_2004}, Cas A \cite{Page:2011wu} and Ioffe 2NT \cite{Yakovlev2nt} for $^{3}\! P_{2}$ neutrons and 
	CCDK \cite{Chen:1993wu} for $^{1}\! S_{0}$ for protons.}
	\label{fig:Tc}
\end{figure}

Of outmost importance in the cooling of neutron stars is the occurrence of pairing, i.e., superfluidity/superconductivity, of baryons
which results from the instability of the Fermi surface under the formation of Cooper pairs in the presence of an attractive interaction 
\cite{Cooper:1956tx,Bardeen:1957tx}.
Various type of pairing are possible depending on the spin and angular momentum of the pairs, the most relevant ones being $^{1}\! S_{0}$,
in spectroscopic notation, corresponding to spin-singlet with orbital angular momentum $L=0$ and total momentum $J=0$
and 
$^{3}\! P_{2}$ corresponding to spin-triplet with $L=1$ and $J=2$.
We refer the reader to the reviews \cite{Page2014,Gezerlis:2014tm} for extensive presentations of the topic.
The first important effect of pairing for our purpose is the appearance of a gap at the Fermi surface in the single particle excitation spectrum, once the temperature
is below the phase transition critical temperature $T_c$, that results in a strong suppression, often as a Boltzmann-like exponential, of
all processes involving the paired component. 
This possibly affects all microphysics ingredients as $c_v$, $K$, and $q_\nu$.
The second important effect is the opening of a new neutrino emission channel in which the energy released in the formation of a Cooper pair
is emitted as a $\nu-\overline{\nu}$ pair.
The pairing phase transition, being of second order, implies the continuous formation and breaking of Cooper pairs as $T$ drops below $T_c$ 
which renders this process, dubbed the ``Cooper Pair Breaking and Formation'', or PBF, very efficient.
For the pairing critical temperature, among the many published models, we choose a few representative ones that we plot in Fig.~\ref{fig:Tc}. In our models, we include the following channels:  $X =\, ^{1}\! S_{0}$ and $Y =\, ^{3}\! P_{2}$ for neutrons, and $Z = \, ^{1}\! S_{0}$ for protons. For clarity in the discussion, we adopt the notation $(X,\,Y,\, Z)$ for labeling the models.

\section{Stellar Structure}
\label{Sec:Structure}

\subsection{Numerical implementation}
\label{Sec:Structure_Numeric}

\begin{figure*}
	\includegraphics[scale=0.60]{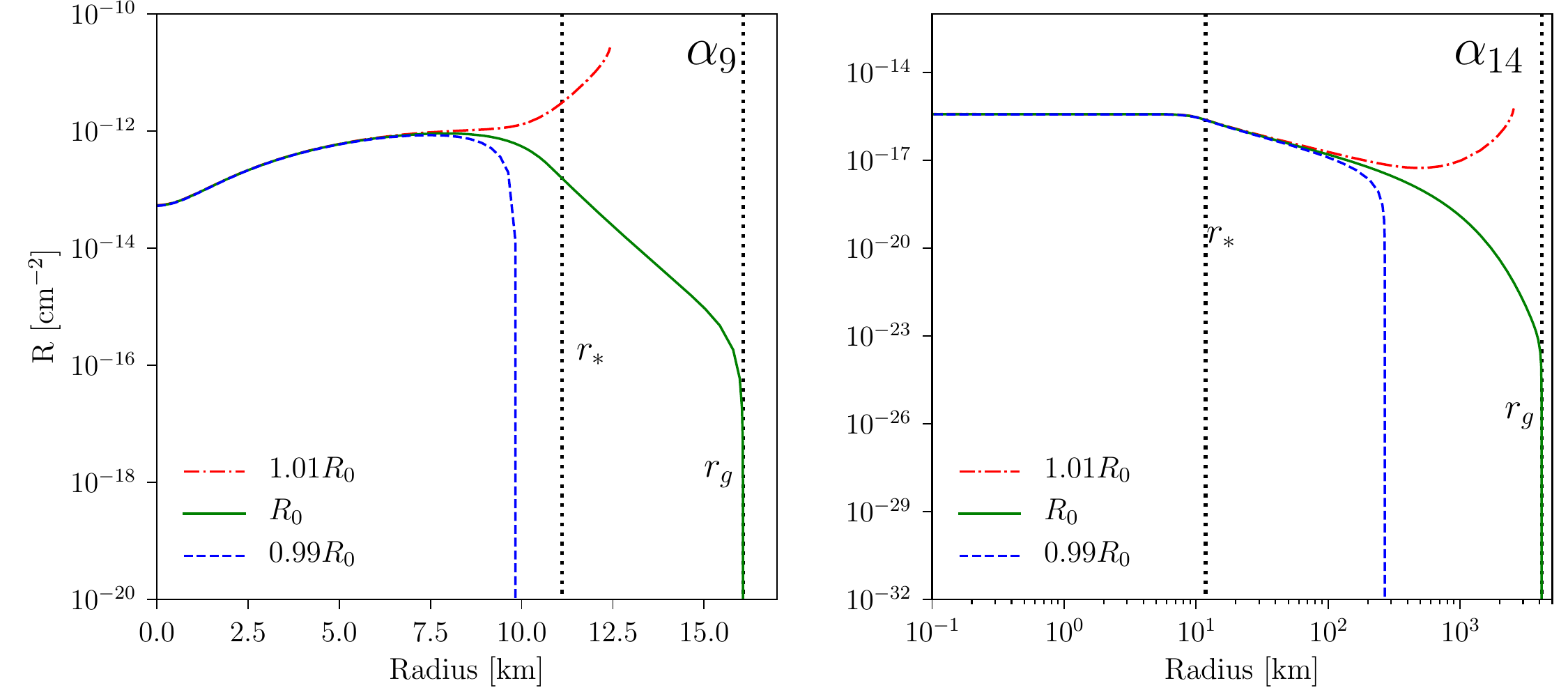}
	\includegraphics[scale=0.60]{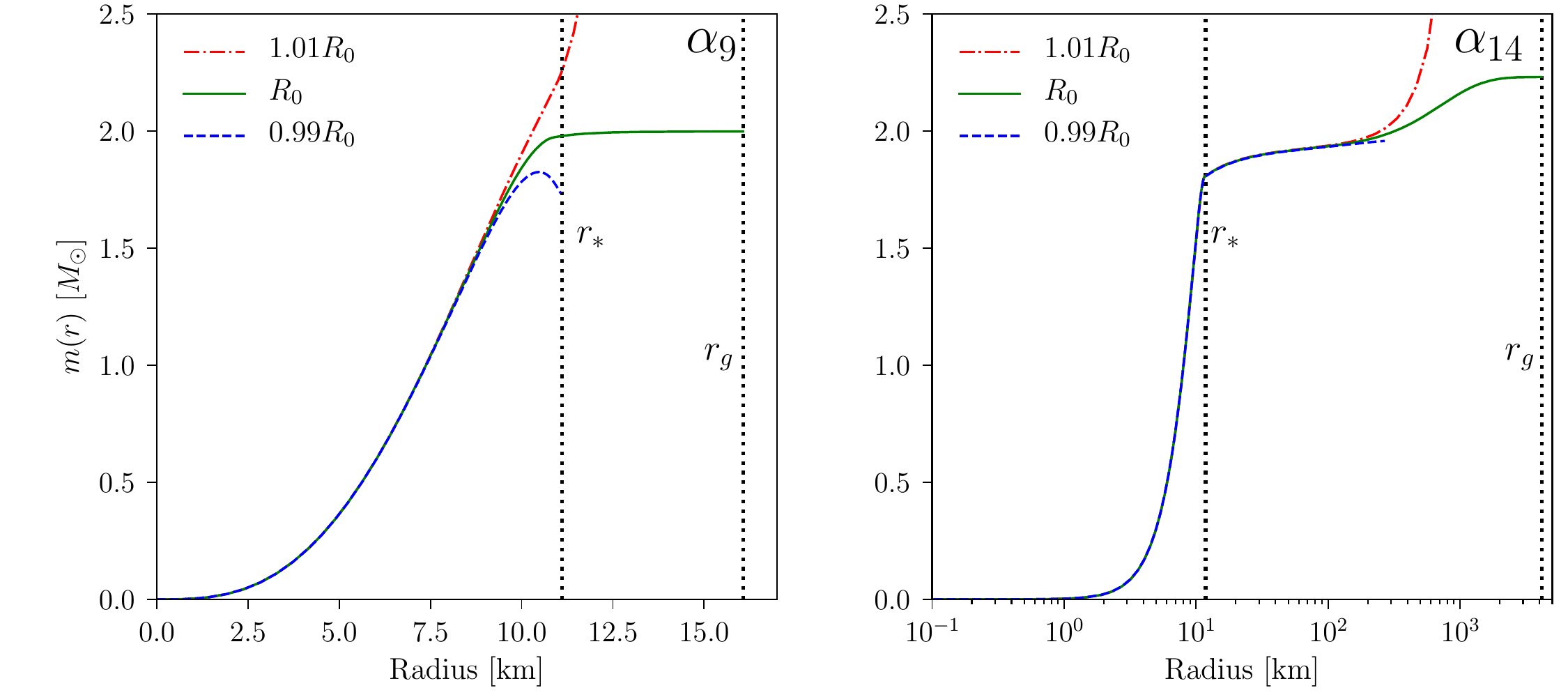}
	\includegraphics[scale=0.60]{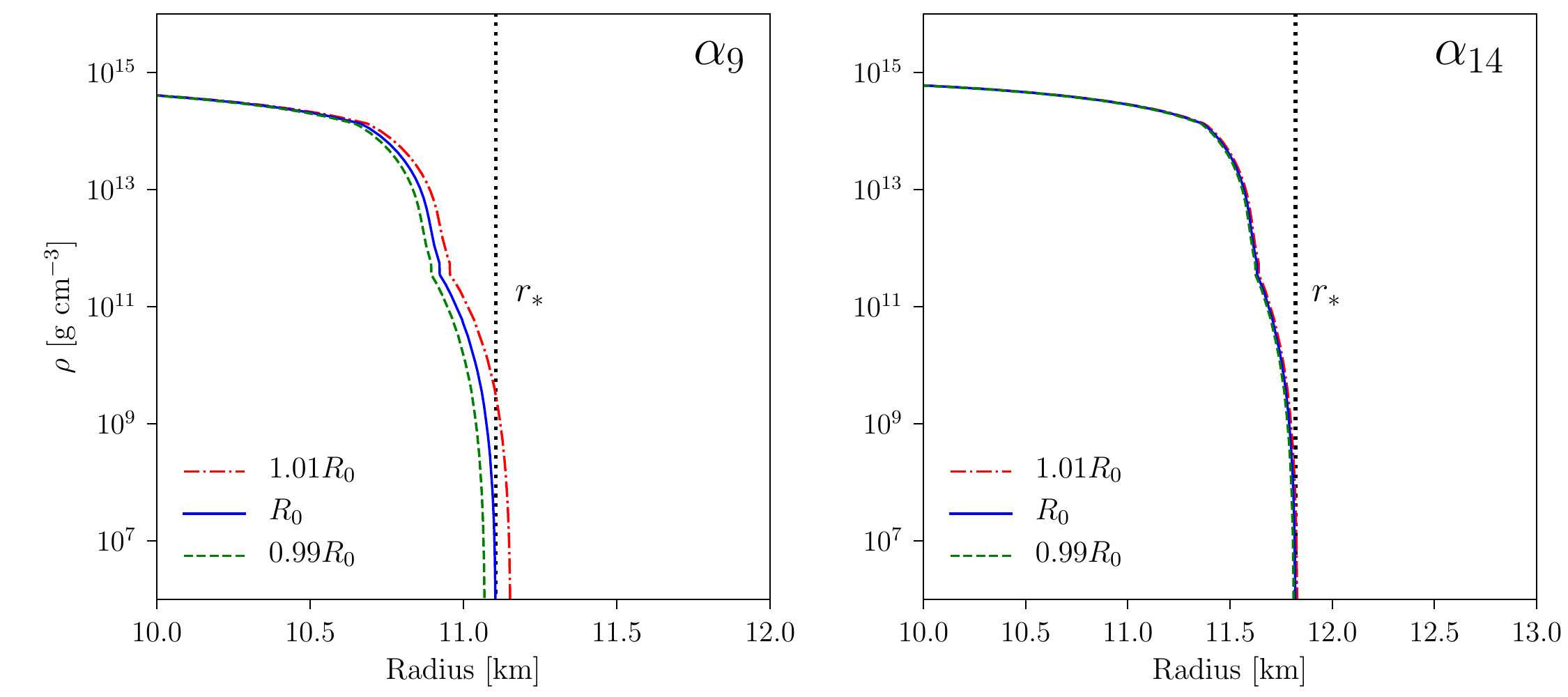}
	\caption{Scalar curvature $R$ (upper panels), mass function $m$ (middle panels) and density $\rho$ in the outer layers (lower panels) 
	for a neutron star model built with the APR EoS
	and central density $\rho_0 = 1.55\times 10^{15}$ g cm$^{-3}$ in $\alpha_{9} R^2$ (left panels) and $\alpha_{14} R^2$ (right panels) gravity
	with the respective locations of their stellar, $r_*$, and gravitational, $r_g$, radii.
	In each panel the three curves show the values corresponding to the central $R_0$ that leads to the Scharzwschild metric beyond $r_g$ and slightly 
	modified values, at $0.99 R_0$ and $1.01 R_0$, that exhibit the divergent behavior of the solutions.}
	 \label{fig:Rrho}
\end{figure*}

\begin{figure*}
	\includegraphics[scale=0.70]{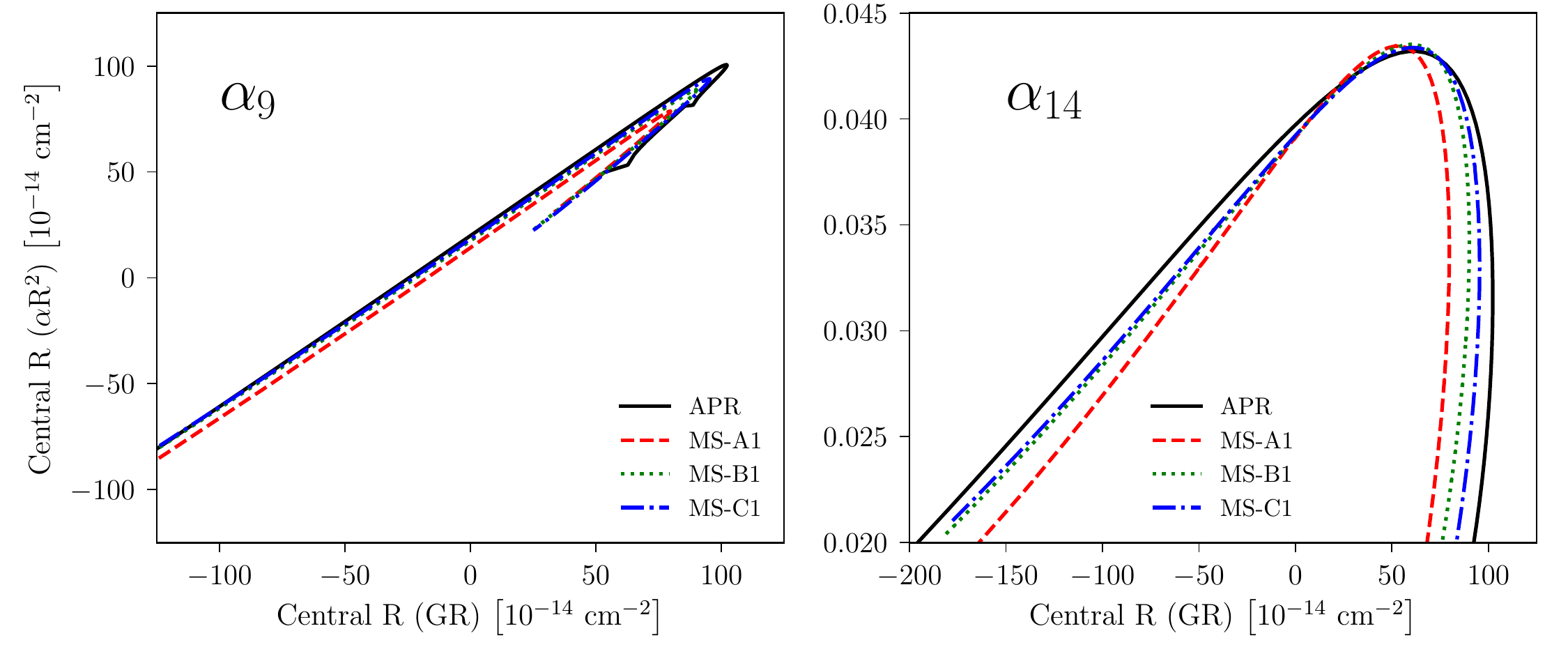}
	\caption{\label{fig:rcenter} Scalar curvature at the star's center for $\alpha_{9} R^2$ (left panel) and $\alpha_{14} R^2$ (right panel) gravity, as a function of its GR counterpart,
	 using four EoSs to build the neutron star models.}
	 \label{fig:RR}
\end{figure*}

The structure equations, (\ref{eq:structure1})-(\ref{eq:structure3}) and (\ref{eq:structure4}),
were solved employing a 4th Order Runge-Kutta method of adaptative stepsize, implemented in FORTRAN77 language. 
The radial interval was split into $[0,r_{\ast}]$ and $[r_{\ast},r_{g}]$, where $r_{\ast}$ is the radius of the star and $r_{g}$ is refered as the gravitational radius. The first is formally defined as the point where $P(r_{\ast})=0$. Numerically, it translates as $P(r_{\ast})\sim\mathcal{O}(10^{-27}P_{0})$, with $P_{0}$ the central pressure. This implies that for $r\in[r_{\ast},r_{g}]$ the system is solved by employing $\varepsilon=P=0$. 
Regarding $r_{g}$, it is defined as the coordinate where $R(r_{g})\sim \mathcal{O}(10^{-30}R_{0})$ beyond which the Schwarzschild metric can be applied.

By demanding regularity of the equations at the origin of coordinates, we obtain the boundary conditions
\begin{equation}
\Lambda(0) = 0,\ \frac{d\Lambda}{dr}\Big|_{r=0}=\frac{d\Phi}{dr}\Big|_{r=0}=\frac{dR}{dr}\Big|_{r=0}=0\ .
\end{equation}
Since $\Phi(r)$ is absent from the right-hand side of the equations, a finite $\Phi(0)$ can be chosen at the beginning of the calculations, and at the end of them perform a re-scaling $\Phi(r)\to\Phi(r)-\Lambda(r_{g})-\Phi(r_{g})$ in order to recover Schwarzschild's solution for $r\geq r_{g}$. 

There is no restriction on the initial condition on the pressure,  $P_0$, which can be chosen arbitrarily, thus generating a family of solutions.
The corresponding energy density, $\varepsilon_{0}$, or mass density, $\rho_0$, are then obtained directly from the EoS. 
In this paper we consider four core EoSs: APR \cite{Akmal:1998vu}, arising from a non-relativistic model with relativistic corrections, 
and three EoSs, MS-A1, MS-B1, MS-C1 which are built from relativistic mean field calculations \cite{Muller:1996wy,Page:2020wj}. 
These are completed by a low density EoS for the inner crust from \cite{Negele:1973tp} and for the outer crust from \cite{Haensel:1989wq}.

Given a central density $\rho_0$, or central pressure $P_0$, to find $R_{0}$ a combination of shooting and bisection methods was implemented.
Two examples of such integrations are shown in Fig.~\ref{fig:Rrho} that illustrate the cause of the numerical subtleties needed to find the correct solution: 
the solution that leads to the Schwarzschild metric at infinity is a bifurcation point.
This figure also shows that for small values of $\alpha$ as $\alpha_9$ the Schwarzschild limit is reached already a few kilometres above the stellar surface while
for extreme cases as $\alpha_{14}$ it is only at distance of several thousands of kilometres that this limit is numerically obtained.
Obtaining a value of the central $R_0$ as close as possible to the exact one is important to describe the metric outside the star however, 
as the lower panels of this figure illustrate, the $\rho$ profile is much more stable that $R$.
In order to overcome these issues, we demanded the bisection method to stop when the length of the interval for seeking $R_{0}$ was inferior to $10^{-27}$ cm$^{-2}$. This stiff condition allowed to obtain variations below 0.05\% for both the final masses $m(r_g)$ and radii $r_{\ast}$, although the stability of the $\rho$ profile guarantees even smaller variations in this stiff scenario.
Taking advantage of an EoS-independent, parabolic-shape relation between $R(0)$ and $R^{\text{GR}}(0)=(8\pi G/c^{4})(\varepsilon_{0}-3P_{0})$, whose order of magnitude is $\alpha$ dependent, as illustrated on Fig.~\ref{fig:rcenter}, it becomes easier to generate families of models when varying $\rho_0$ once a first solution has been found.

Finally, it is important to notice that given the values of $R$ found after integrations, for the whole range of $\alpha$ we have explored and as illustrated in Fig.~\ref{fig:rcenter},
$\alpha R$ is always smaller than $10^{-2}$, i.e., the term $\alpha R^2$ always results to be a small correction to $R$ in the Lagrangian density $\cal L$.
Increasing values of $\alpha$ actually force smaller values of $R$ when the structure equations are solved:
while $R^{(\mathrm{GR})}$ is always $\sim 10^{-12}$ cm$^{-2}$ in the center of a neutron star, in $\alpha R^2$ gravity $R_0$ is significantly reduced.
A consequence of this, as we will show below, is that when values of $\alpha$ beyond $10^{12}$ cm$^2$ are considered the resulting structure of a neutron star
remain almost identical to the $\alpha_{12}$ case, i.e., there is a clear saturation of the effects of $\alpha R^2$ gravity at large $\alpha$. 
However, when considering the shape of the metric outside the star, increasing values of $\alpha$ beyond $\alpha_{12}$ still imply increasing values of $r_g$.

\subsection{Gravitational and baryonic masses}
\label{sec:mass}

\begin{figure*}[t]
	\includegraphics[scale=0.6]{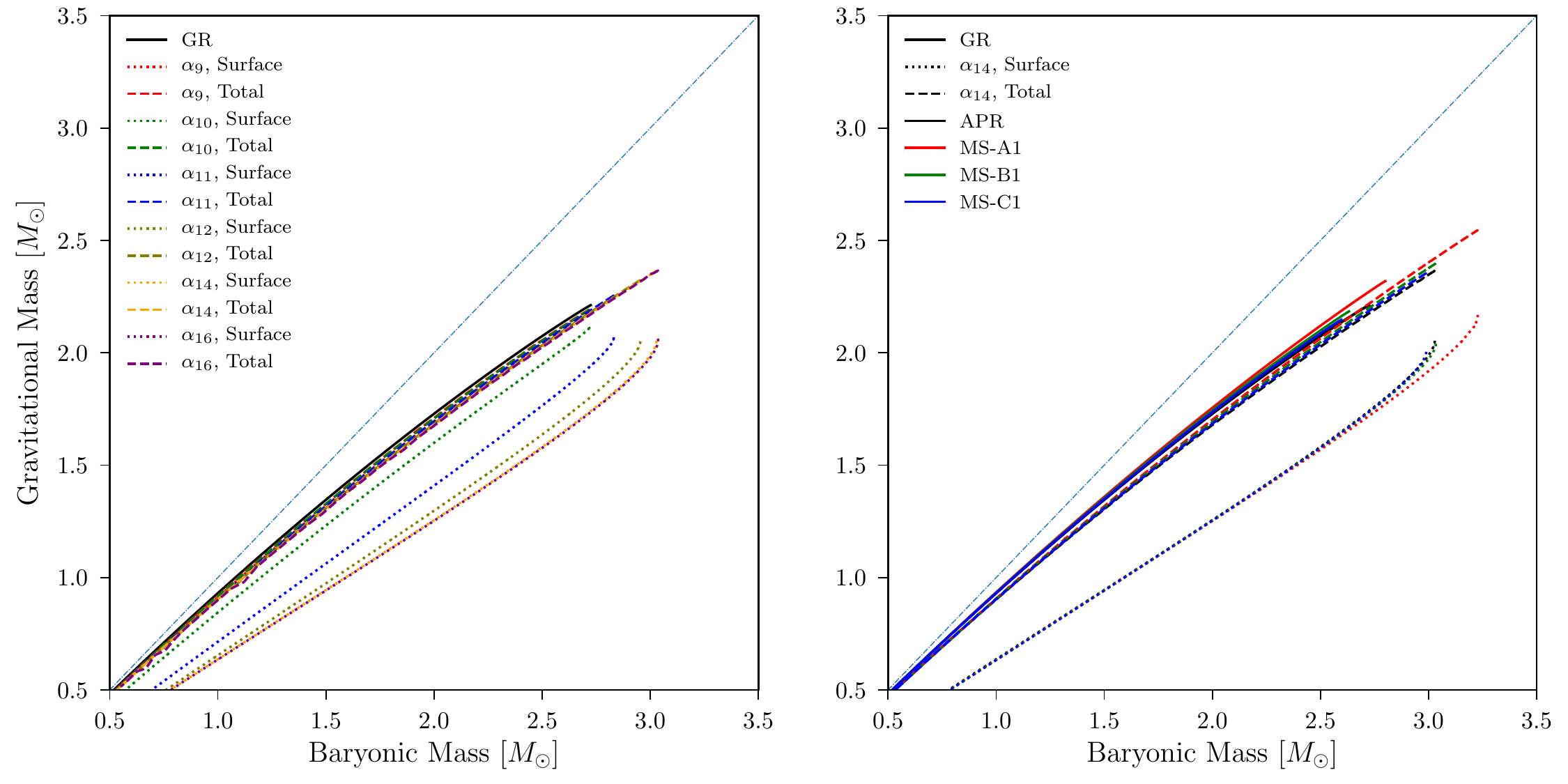}
	\caption{Gravitational mass in GR and total gravitational and surface masses in $\alpha_{14}R^2$ gravity, versus baryonic mass.
	(We also plot the diagonal dash-dot curve to guide the eye.)}
	\label{fig:mbarionic}
\end{figure*}

\begin{figure*}
	\includegraphics[scale=0.6]{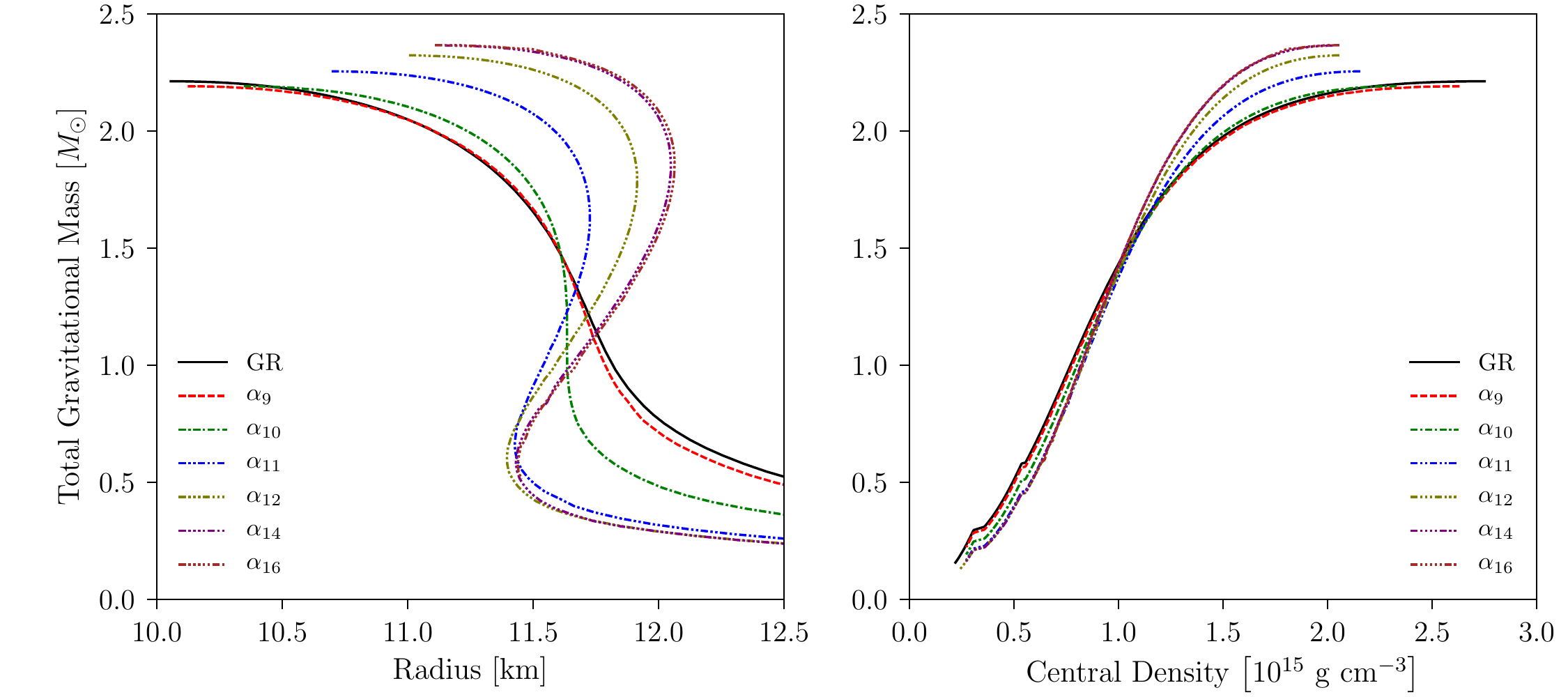}
	\caption{ 
		Total gravitational mass versus: (left panel) radius; (right panel) central density, considering the APR EoS and 
		both GR and $\alpha R^2$ theories and various values of $\alpha$ as labeled.
		}
		\label{fig:mreosa}
\end{figure*}

\begin{figure*}
	\includegraphics[scale=0.6]{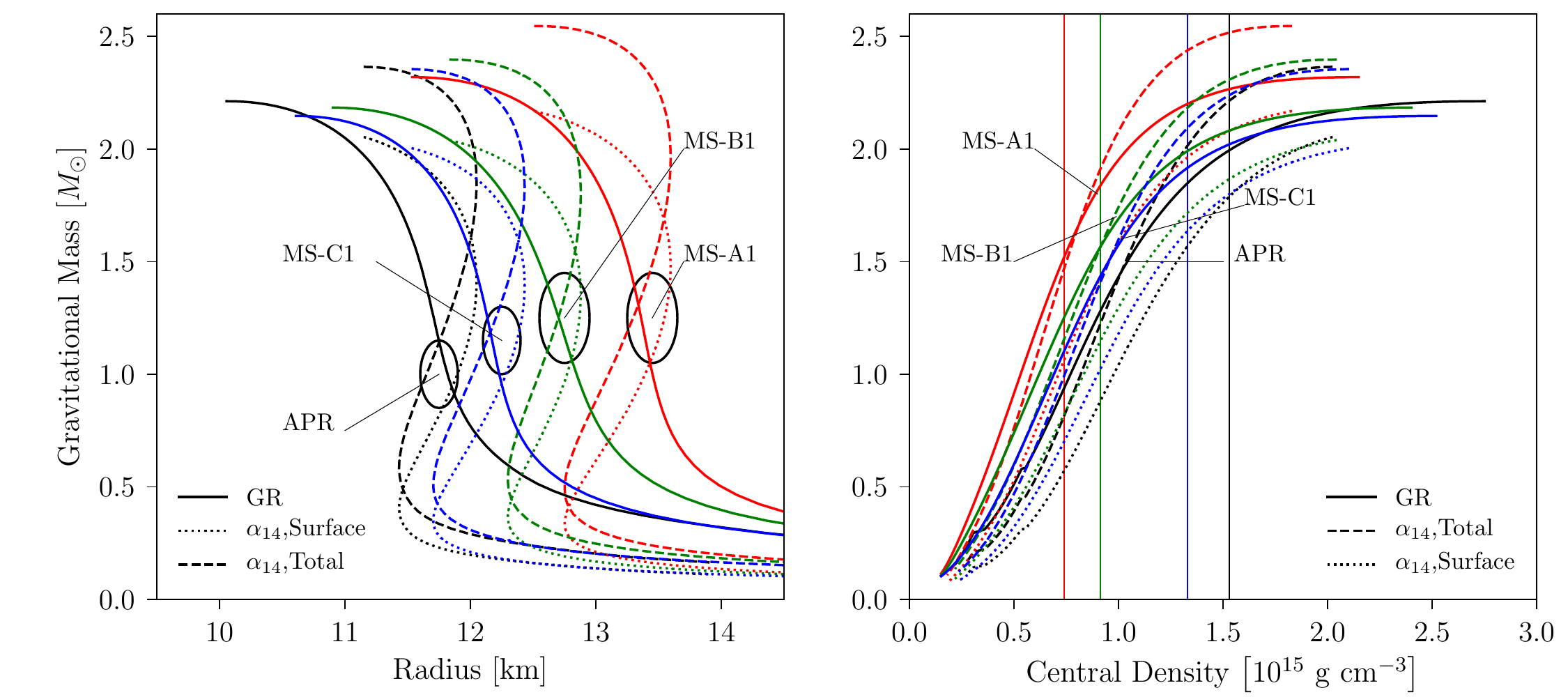}
	\caption{
	Gravitational mass for the labeled EoSs, GR and $\alpha_{14}R^2$ theories, against: (left) radius; (right) central density.
	The four vertical lines in the right panel mark the DUrca critical densities, $\rho_\text{DU}$, for MS-A1, MS-B1, MS-C1 and APR,
	by order of increasing density.
	}
	\label{fig:mreosb} 
\end{figure*}

In analogy to the Schwarzschild metric, we can introduce a mass function, $m(r)$, by writing
\begin{equation}
\exp(2\Lambda) = \frac{1}{1-2Gm(r)/rc^2} \quad .
\label{eq:mr}
\end{equation}
Under the condition that we recover the Schwarzschild metric in the limit $r\to\infty$ we obtain the \textit{total} gravitational mass of the star as
\begin{equation}
M_\text{tot} := \lim_{r\to\infty}m(r) \quad .
\end{equation}
In practice we use $M_\text{tot} = m(r=r_g)$.
We will also consider a \textit{surface} gravitational mass as $M_\mathrm{surf}=m(r=r_{\ast})$.
Another mass concept frequently encountered is the \textit{baryonic} mass, $M_\mathrm{bar}$, 
defined through the total baryon number of the star, $N_\mathrm{bar}$, as
\begin{equation}
M_\mathrm{bar} \equiv m_B N_\mathrm{bar} = m_B \int^{r_*}_{0}4\pi r^{2} \ n(r) \, e^{\Lambda(r)} \, dr\ ,
\end{equation}
where the $e^{\Lambda(r)}$ factor corrects for the proper radial length from the areal radius $r$, 
and $m_B$ is the "baryon mass" taken as the neutron, or proton, or atomic unit, mass.
These concepts are necessary to understand both the cooling setup and results: 
the surface gravitational mass is the value we shall employ to define the surface gravity (see Eq.~\ref{eq:gsdefin2} below), 
while the total gravitational mass is the value measured in the case of binary systems.
The baryonic mass is just a reference mass and is not a measurable quantity but is a useful conserved value in some contexts.

For any given EoS, and a chosen gravity theory, we can calculate a family of neutron star models by varying the central density $\rho_0$.
With such results, for the APR EoS, we compare in the left panel of Fig.~\ref{fig:mbarionic} the gravitational mass in GR and the total and surface gravitational masses
in $\alpha R^2$ gravity for various values of $\alpha$ as function of the baryonic mass.
We see that $M_\mathrm{tot}$ is slightly lower while $M_\mathrm{surf}$ is significantly lower than the GR gravitational mass, all of them being lower than $M_\mathrm{bar}$.
A very interesting result is that in the observed range of gravitational masses, between 1.2 up to 2.2 $M_\odot$,
 the difference between $M_\mathrm{tot}$ and $M_\mathrm{surf}$ is almost independent of $M_\mathrm{bar}$ and is a strong function of $\alpha$.
In the right panel of the same figure we compare the cases of GR and $\alpha_{14} R^2$ gravity but employing our four EoSs and see
that this mass difference in almost independent of the EoS.

With the same procedure of varying $\rho_0$ we can generate $M-r_{\ast}$ and $M-\rho_0$ curves which we present in Fig.~\ref{fig:mreosa} and ~\ref{fig:mreosb}.
In agreement with previous works on the subject, we find that $M_\text{tot}$ can reach larger values in the $\alpha R^{2}$ theory than in GR.
The left panel of Fig.~\ref{fig:mreosa} shows that there is always a crossing of the $M-r_{\ast}$ curves of GR and any $\alpha R^{2}$ theory and 
that at masses below the crossing point $\alpha R^{2}$ NS models have smaller radii while above this point they have larger radii than GR models of the same total gravitational mass.
(For small values of $\alpha$, as $\alpha_{9}$ and $\alpha_{10}$, there appears a second crossing point a $M$ close to the maximum mass.)
On the right panel of Fig.~\ref{fig:mreosa} we also find a crossing point on the $M-\rho_0$ curves:
significant differences in the predicted total gravitational masses, for a given $\rho_0$, between GR and $\alpha R^{2}$ theories with large $\alpha$ 
appear above this crossing point while at lower central densities the differences are much smaller.
We can also notice that the central density for the maximum mass is progressively lower as $\alpha_X$ grows in the case of the APR EoS.

These considerations are also valid in the case of the other three EoSs we consider as shown in Fig.~\ref{fig:mreosb}.
Finally, we see that $\alpha_{14}$ and $\alpha_{16}$ results are almost identical and illustrate the saturation of the $\alpha R^2$ effects
of which we will find more examples below.

\subsection{Stellar proper volume}

\begin{figure}
	\includegraphics[scale=0.6]{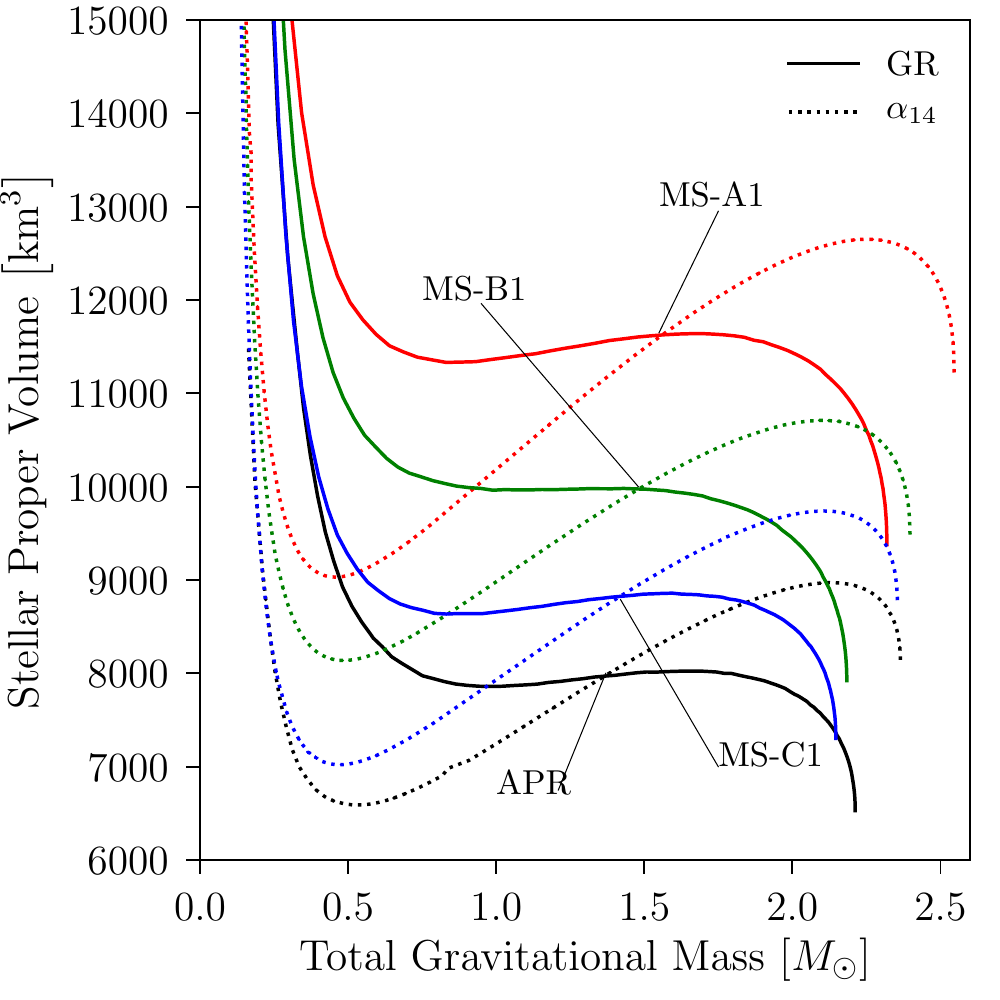}
	\caption{Total stellar volume, considering the four EoSs, in GR versus $\alpha_{14}R^2$ gravity.}
	\label{fig:gsduvapr}
\end{figure}

The proper volume of a star is naturally defined as
\begin{equation}
V = \int^{r_*}_{0}4\pi r^{2} \, e^{\Lambda(r)} \, dr\ ,
\end{equation}
where the $e^{\Lambda(r)}$ factor corrects for the proper radial length from the areal radius $r$.
On Fig.~\ref{fig:gsduvapr}, the stellar proper volume of the objects is plotted against the total gravitational mass. 
There is a clear EoS-independent trend that at high masses, higher than about $1.4-1.5 \; M_\odot$ depending on the EoS, the proper volume of the 
star in $\alpha R^2$ gravity is larger than in GR for $\alpha_{14}$ while at lower masses the effect is inverted.
This effect parallels the relative variation in radii exemplified in the left panel of Fig.~\ref{fig:mreosb}.

\subsection{The DUrca critical mass and volume}

While it is clear that $\rho_\text{DU}$ is independent of the gravity theory, depending only on the EoS, the critical stellar mass $M_\text{DU}$
at which the DUrca process becomes allowed does depend on the gravity theory.
Specifically, the respective values of  $\rho_\text{DU}$ are $7.41 \times 10^{14}$, $9.13\times 10^{14}$, $1.33\times 10^{15}$ and $1.53\times 10^{15}$ 
g cm$^{-3}$ for the MS-A1, MS-B1, MS-C1 and APR EoS, respectively.
Similarly, the proper volume of the inner core where the DUrca process is allowed is of importance.
This quantity is defined as
\begin{equation}
V_{\text{DU}} = \int^{r_{\text{DU}}}_{0}4\pi r^{2}e^{\Lambda(r)}dr\ ,
\end{equation}
with $r_{\text{DU}}$ the last radial coordinate where the criterium of Eq.~\ref{eq:DU} is satisfied, i.e., where $\rho_\text{DU}$ is reached.
 
\begin{figure*}
	\includegraphics[scale=0.6]{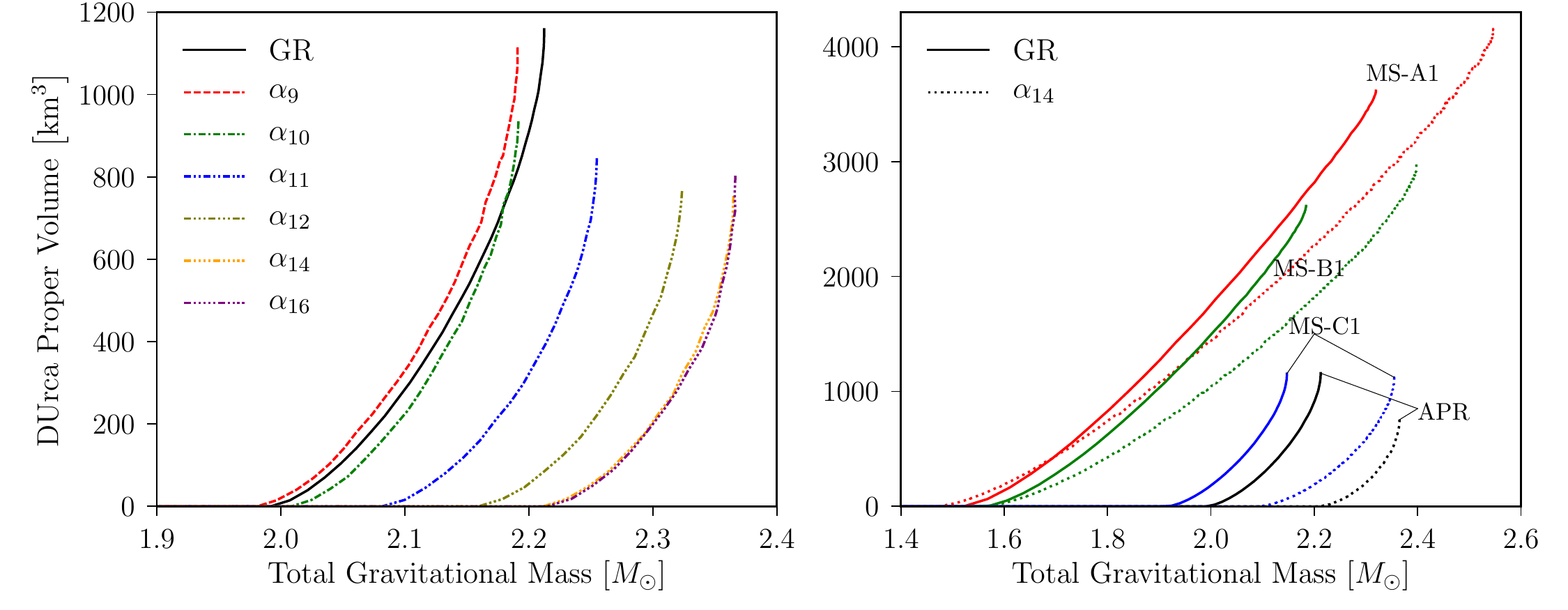}
	\caption{Proper volume of DUrca process against total gravitational mass in GR versus $\alpha R^2$ gravity for:
	(left panel) APR EoS and the indicated values of $\alpha_{X}$; (right panel) several EoSs, GR and $\alpha_{14}$.}
	\label{fig:duveos} 
\end{figure*}

In the left panel of Fig.~\ref{fig:duveos} we show how $V_{\text{DU}}$ grows with $M$ for the APR EoS in GR and in 6 $\alpha R^2$ theories.
One clearly sees that the threshold mass $M_\text{DU}$ increases significantly as $\alpha$ is increasing while the maximum $V_\text{DU}$,
reached at the maximum mass, decreases.
When considering other EoSs, in the right panel of Fig.~\ref{fig:duveos}, the same behavior, increase of $M_\text{DU}$ with $\alpha$, 
is seen for the EoS MS-C1 while for MS-B1 there is little change in $M_\text{DU}$ and for MS-A1 the contrary happens, $M_\text{DU}$ decreases with $\alpha$.
The reason for this difference is simple when one considers the crossing point of the $M-\rho_0$
curves seen in the right panels of Fig.~\ref{fig:mreosa} and \ref{fig:mreosb}: 
$\rho_\text{DU}$ for MS-A1 and MS-B1 is close to the crossing point, i.e., in the density regime where the $M-\rho_0$ curve is little affected by $\alpha$
while for the EoSs MS-C1 and APR, $\rho_\text{DU}$ is much higher than the crossing point and changes in $\alpha$ have a large impact.

\subsection{Crust thickness}

\begin{figure*}
	\includegraphics[scale=0.6]{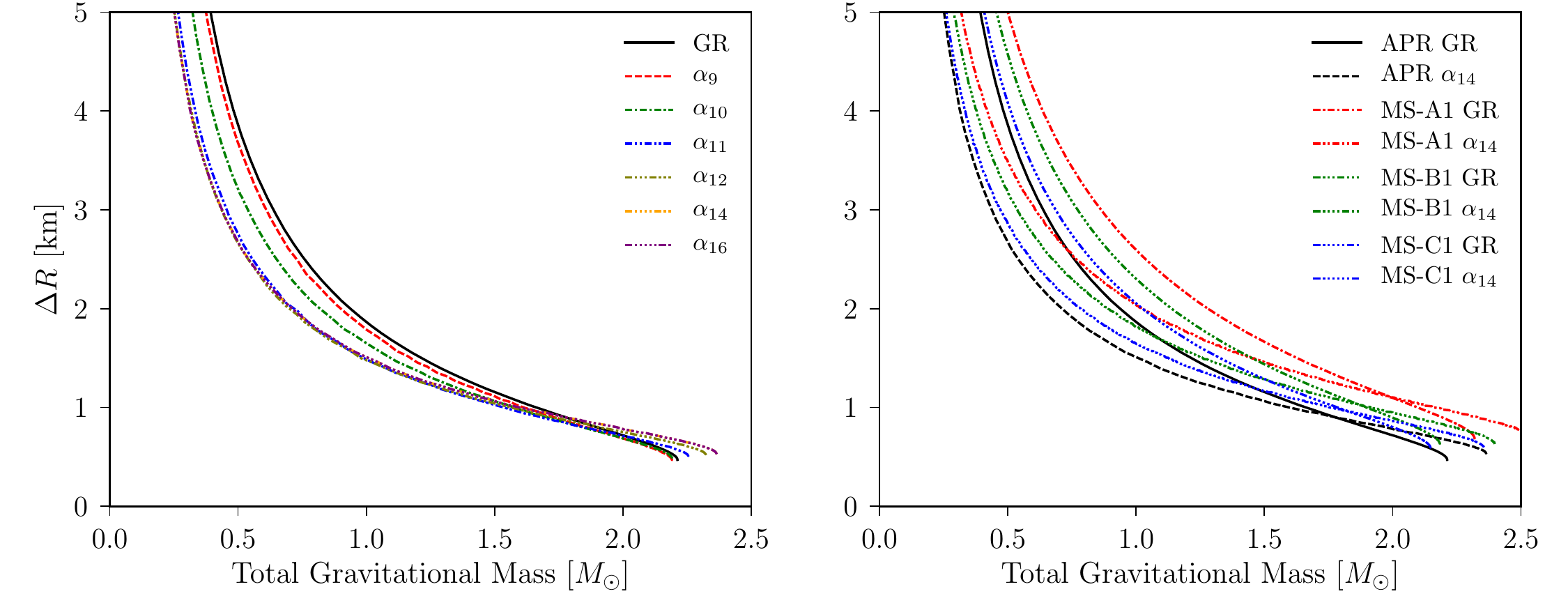}
	\caption{Crust thickness as a function of the total gravitational mass, in GR versus $\alpha R^2$ gravity for: 
	(left panel) APR EoS and several values of $\alpha_{X}$; 
	(right panel) the indicated EoSs, GR and $\alpha_{14}$.}
	\label{fig:crustt}
\end{figure*}

Let $r_{cc}$ be the radial coordinate where the crust-core transition takes place, at a density $\rho_{cc} = $1.4$\times 10^{14}$ g cm$^{-3}$ \cite{Lorenz:1993tz}.
The crust thickness is defined by
\begin{equation}
\Delta R = \int^{r_{\ast}}_{r_{cc}}e^{\Lambda(r)}dr
\end{equation}
with again an $e^{\Lambda(r)}$ factor to correct from the areal radius to the proper radius.
On the left panel of Fig.~\ref{fig:crustt} we can notice that generally the thickness gradually diminishes as $\alpha$ increases while it slightly increases
at $M \gtrsim 1.8 M_\odot$.
This reduction in $\Delta R$ is very significant at low masses while in the relevant mass range $1.2 - 2.2 M_\odot$ changes in $\Delta R$ are effectively minor.
It is interesting to notice that at $0.5 M_\odot$ the radius of the star shrinks by 1 km when comparing GR and $\alpha_{14}$ models (see Fig.~\ref{fig:mreosa})
which appears to be almost totally due to the shrinking of the crust that is also of the order of 1km.
In contradistinction, at larger masses the change in $R$ is much larger that the change in $\Delta R$, i.e., most of the changes are occurring in the core. 
The right panel of Fig.~\ref{fig:crustt} sows that these effect are quite independent of the employed EoS.

\subsection{Surface gravity}
\label{subsec:surfgrav}

\begin{figure*}
	\includegraphics[width=0.8\textwidth]{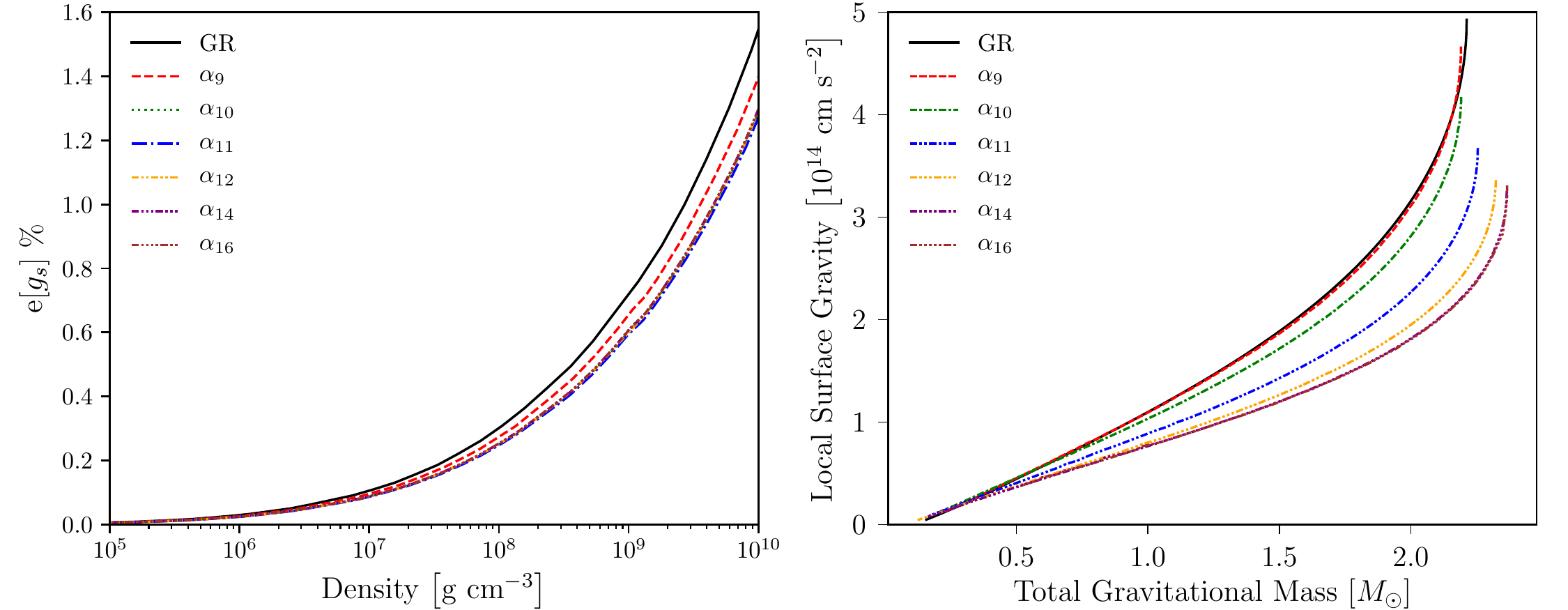}
	\caption{Gravitational acceleration at the surface of neutron star models built with GR versus $\alpha R^2$ gravity.
	Left panel: percentual difference between $g_{s}$ and $g(r)$ at densities close to the surface of the star, considering an APR model with 
	$\rho_{0}=1.5\times 10^{15}$ g cm$^{-3}$. Right panel: local surface gravity for the APR EoS and the indicated values of $\alpha_{X}$.}
	\label{fig:gserr}
\end{figure*}

In General Relativity, near the surface of the star the hydrostatic equilibrium equation reduces to
\begin{equation}
\frac{dP}{dr} = -g_{s}\rho\ ,
\label{eq:dPdr}
\end{equation}
where $g_{s}$ is the \textit{local surface gravity} defined by
\begin{equation}
g_{s}:=g(r)\Big|_{r=r_{\ast}}
\label{eq:gsdefin}
\end{equation}
with
\begin{equation}
g(r) = c^{2}e^{-\Lambda(r)}\frac{d\Phi}{dr}\ .
\label{eq:gsdefin1}
\end{equation}
We, thus, obtain a simple expression for the surface gravity in GR as
\begin{equation}
g_{s} = \frac{GM}{r^{2}_{\ast}}\left[1-\frac{2GM}{c^{2}r_{\ast}}\right]^{-1/2}\ .
\label{eq:gsdefin2}
\end{equation}
with $M$ the gravitational mass. 
Within the $\alpha R^{2}$ theory it is trivial to verify that Eq.~\ref{eq:dPdr}, \ref{eq:gsdefin} , and \ref{eq:gsdefin1}, also hold.
To verify that Eq.~\ref{eq:gsdefin2} holds let us define 
\begin{equation}
e\left[g_{s}\right]\% = 100\times\Bigg|\frac{g_{s}-g(r)}{g_{s}}\Bigg|
\end{equation}
where $g_s$ is calculated from Eq.~\ref{eq:gsdefin2} with $M \rightarrow M_\mathrm{surf}$ and $g(r)$ from Eq.~\ref{eq:gsdefin1}.
This quantity is plotted in the left panel of Fig.~\ref{fig:gserr} which proves that employing Eq.~\ref{eq:gsdefin2} for $g_s$ is accurate to within 1\%
within $\alpha R^2$ models as well as in GR.
It is, hence, acceptable to keep $g_{s}$ as defined above, with $M=M_\mathrm{surf}$.  
On the right panel of Fig.~\ref{fig:gserr} it is clear that $\alpha R^{2}$-stars have lower values of $g_{s}$ in comparison with their GR counterparts
for the same (total) gravitational mass,
a direct effect of reduction of $M_\mathrm{surf}$ with respect to $M_\mathrm{tot}$.

\section{Thermal Evolution}
\label{Sec:Cooling}

\subsection{Numerical implementation and envelope models}
\label{Sec:Cooling_Numeric}

\begin{figure}
	\includegraphics[scale=0.27]{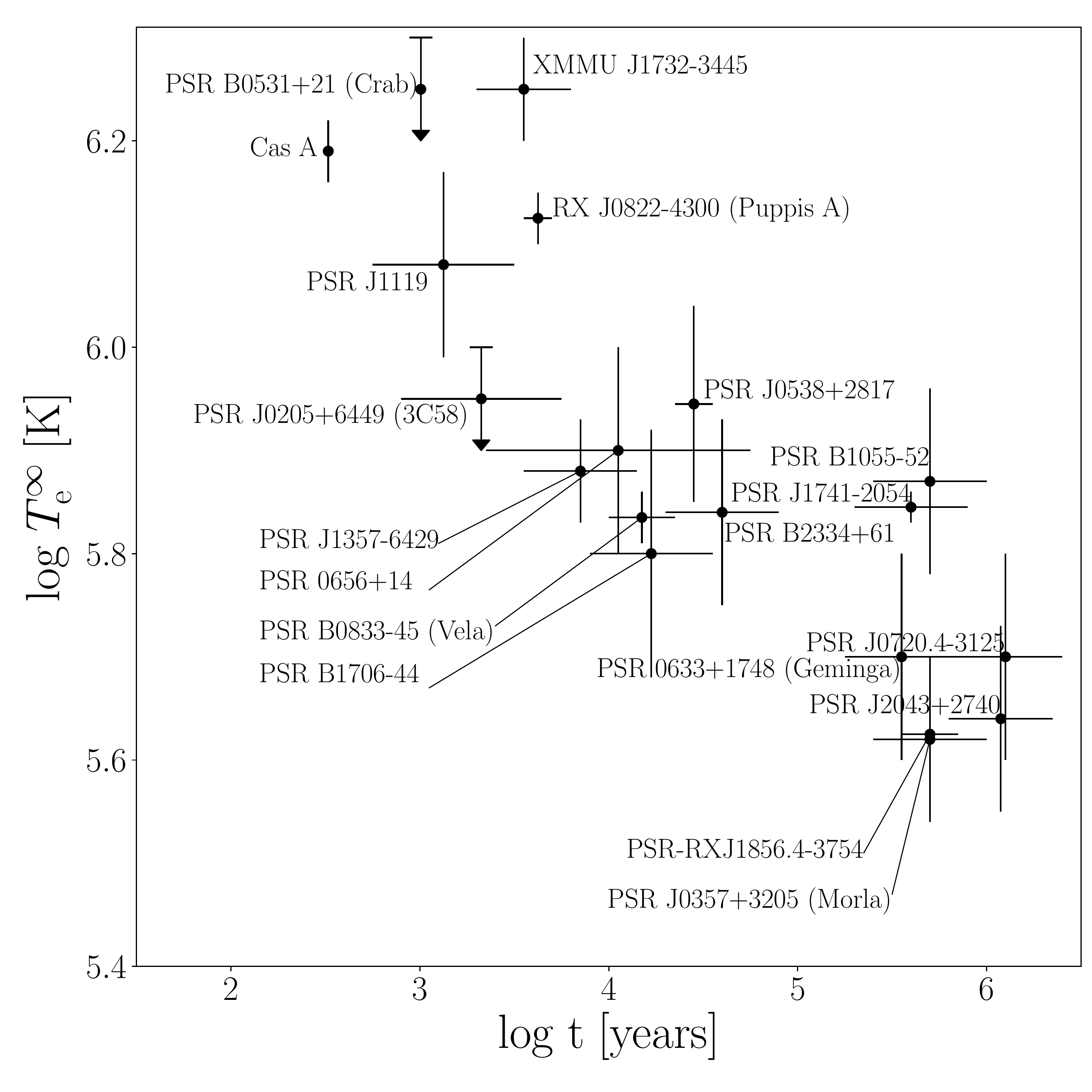}
	\caption{\label{fig:odata} The sample of observational data considered for this work, taken from \cite{Beznogov:2015wn}.
	The name of each object is given, and the error bars correspond to 1$\sigma$ .
	}
\end{figure}

For simulating the thermal evolution of NSs, the fully relativistic 1-D code \texttt{NSCool} was employed \cite{Page:2016vf}. 
As is traditional in neutron star cooling modeling, the star is split in two parts:
the interior where the EoS is treated as temperature independent, and a low density outer region, the envelope, where the EoS is $T$-dependent.
In the interior the equations \ref{Eq:dTdr_GR} and ~\ref{Eq:Ebalance_GR} are solved with a fully implicit Heyney scheme with the inner boundary condition
\begin{equation}
L(r=0) = 0
\end{equation}
and an outer boundary condition, at a density $\rho_b$, provided by an envelope model.
In such a scheme the hydrostatic structure of the interior does not change with time and is only calculated once at the beginning of the simulation. 
For the core EoS we consider the same 4 models as in the previous section.

In an envelope model for the outer boundary condition it is assumed that 
1) there are no significant energy sources or sinks within it and
2) the thermal time-scale of this thin layer is much shorter than the one of the interior. 
Thus, the time derivatives in Eq.~\ref{Eq:Ebalance_GR} are neglected and,
as a result, in the envelope the luminosity is uniform, i.e. $L(r) = L_*$ where $L_*$ is the surface luminosity, and only the heat transport equation ~\ref{Eq:dTdr_GR}
needs to be solved.
The integration starts with choosing a surface or {\it effective} temperature, $T_e$, and define the luminosity as
\begin{equation}
L_* = 4 \pi r_*^2 \sigma_\mathrm{SB} T_e^4
\label{eq:LTe}
\end{equation}
in analogy with the blackbody flux $F=\sigma_\mathrm{SB} T^4$, $\sigma_\mathrm{SB}$ being the Stefan-Boltzmann constant.
Using a photosphere condition (typically the Eddington condition, see, e.g., \cite{Kippenhahn:2012tm}) the density $\rho_e$ and pressure $P_e$
corresponding to $T_e$ are obtained and from this Eq.~\ref{Eq:dTdr_GR},
and the simplified version of the hydrostatic equilibrium \ref{eq:dPdr} adequate for a thin envelope,
are integrated inward until $\rho$ reaches $\rho_b$ with a temperature $T_b$.
Notice that, once taken into account the trivial adjustment of $g_s$ in Eq.~\ref{eq:gsdefin2}, the equations to be solved, Eq.~\ref{Eq:dTdr_GR} and  \ref{eq:dPdr},
are the same as in GR and thus all envelope models calculated within GR automatically apply within $\alpha R^2$ gravity.
For a given $g_s$, by varying $T_e$ one gets a family of models generating what is commonly referred to as a ``$T_e - T_b$ relationship'', 
$T_e = T_e(T_b, g_s)$, which we take as our outer boundary condition:
at $\rho_b$ the temperature $T_b$ and luminosity $L_b$ must be such that 
\begin{equation}
L_b = L_* = 4\pi r_*^2 \sigma_\mathrm{SB} T_e(T_b, g_s)^4 \quad .
\label{eq:Lbund}
\end{equation}
It turns out there is a simple scaling of $T_e$ in terms of $g_s$ \cite{Gudmundsson:1983wm} as $T_e(T_b,g_s) = g_{s 14}^{1/4} T_e(T_b, g_{s 14} = 1)$
where $g_{s 14} = g_s/10^{14}$ cm s$^{-2}$.
This envelope acts as a thermal insulator between the hot interior and the surface and a lower $g_s$ allows for a thicker envelope, hence a more efficient
insulation, resulting in a lower $T_e$, for a given $T_b$, as described by this scaling.

 Various types of envelope models have been presented considering variations in the strength and geometry of 
the magnetic field \cite{Page:1996ur} and the chemical composition \cite{Potekhin:1997mn}.
It turns out that, excluding the case of magnetar size magnetic fields, the effects of the chemical composition are much more significant
than the magnetic ones \cite{Potekhin:2003wi} and we will here neglect the latter.
Variation in the chemical composition range from the presence of low-$Z$ elements such as $^{4}$He and $^{12}$C up to iron-like ones.
Low $Z$ material having higher thermal conductivity that high $Z$ ones, their presence in the envelope will result in an increase of $T_e$, for a given $T_b$,
compared to the case of an envelope made of the latter.
This effect is accounted in the $T_e-T_b$ relationship through the adimensional parameter
\begin{equation}
\eta = g^{2}_{s,14}\frac{\Delta M_L}{M_\mathrm{surf}} = \left(\frac{P_L}{1.193\times 10^{34}\ \text{dyn\, cm}^{-2}}\right)\ ,
\label{eq:eta}
\end{equation}
where $\Delta M_L$ is the mass of this light element layer, $M_\mathrm{surf}$ is the surface gravitational mass of the star, and $P_L$ is the maximum pressure at which low-$Z$ exist.
When the outer layers of a newborn neutron star are formed at very high temperatures it is natural to expect that matter will reach its ground state,
commonly called ``catalyzed matter'', that consist of iron-peak nuclei (see, e.g., section 3.1 in \cite{Haensel:2007un}).
However late fall-back after the supernova explosion of posterior accretion may, or may not, deposit light elements at the surface and this results in a large
uncertainty in $\eta$ that is an inherent aspect of the cooling theory.
There is to date, unfortunately, no known way of determining the thickness of this light element layer.

To contrast theoretical predictions with observational data, we employ the sample of \cite{Beznogov:2015wn}, consisting of 19 objects for which the effective temperature and age have been constrained, as illustrated in Fig.~\ref{fig:odata}.

In the following we begin with simple cooling models that emphasize the difference between $M_\mathrm{tot}$ and $M_\mathrm{surf}$
while the next subsection consider more realistic models that include pairing and uncertainties in the outer layers, the envelope, structure.
The last two subsections focus on two particular stars which are the two youngest known neutron stars and present interesting features.

\subsection{Simple cooling models and impact of local surface gravity}
\label{sec:cooling_simple}

\begin{figure*}
	\includegraphics[scale=0.6]{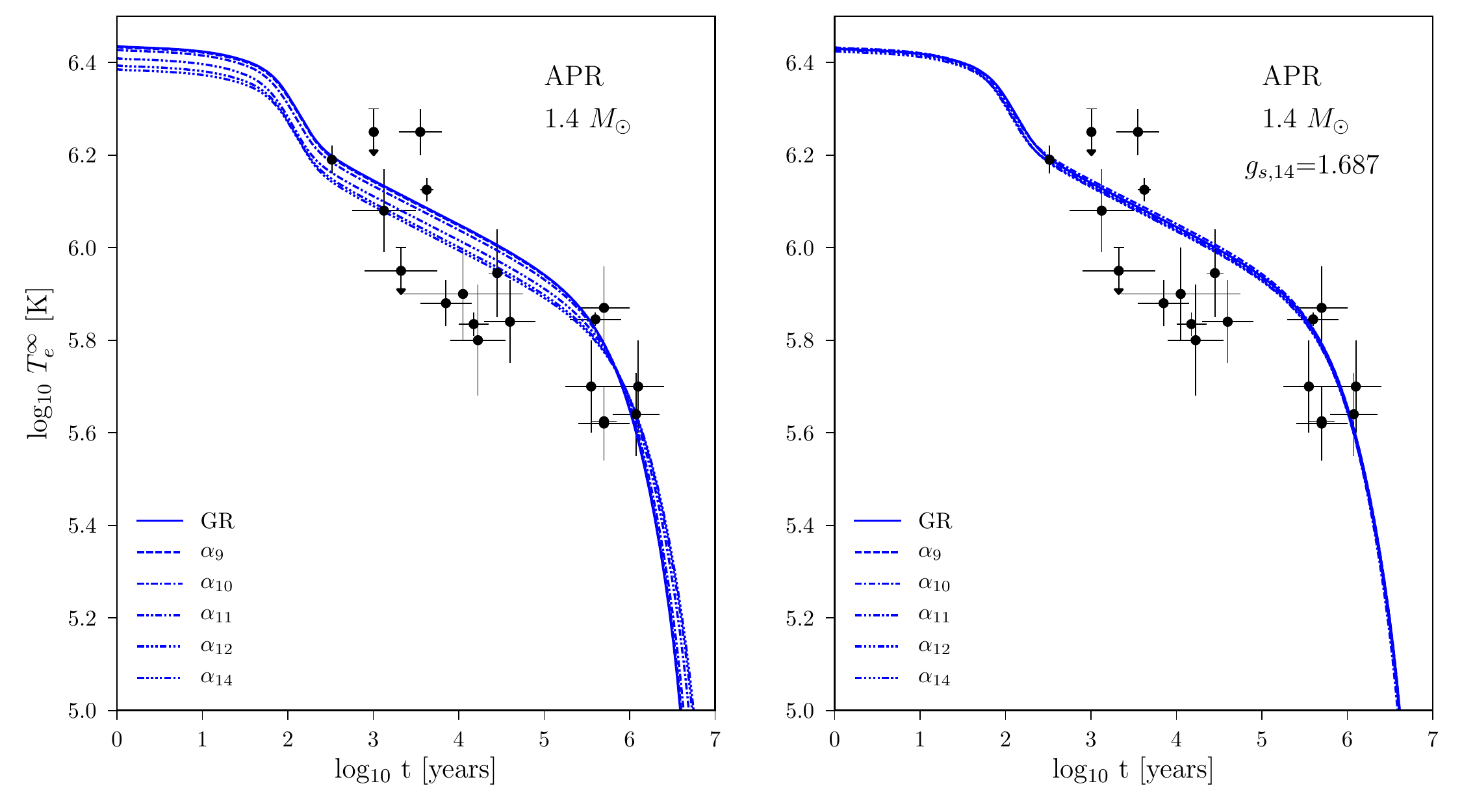}
	\caption{ 
	Left panel: cooling curves of a simple 1.4$M_{\odot}$ APR-NS model, within GR and $\alpha_X R^2$ gravity with $X=9-14$. 
	Right panel: same set of models but with the surface gravity $g_{s}$ fixed to the GR model's value, 1.687$\times 10^{14}$ cm s$^{-2}$.
        In both panels superfluidity/superconductivity is absent and the envelope is made of heavy elements.}
        \label{fig:cool1}
\end{figure*}

\begin{figure*}
	\includegraphics[scale=0.6]{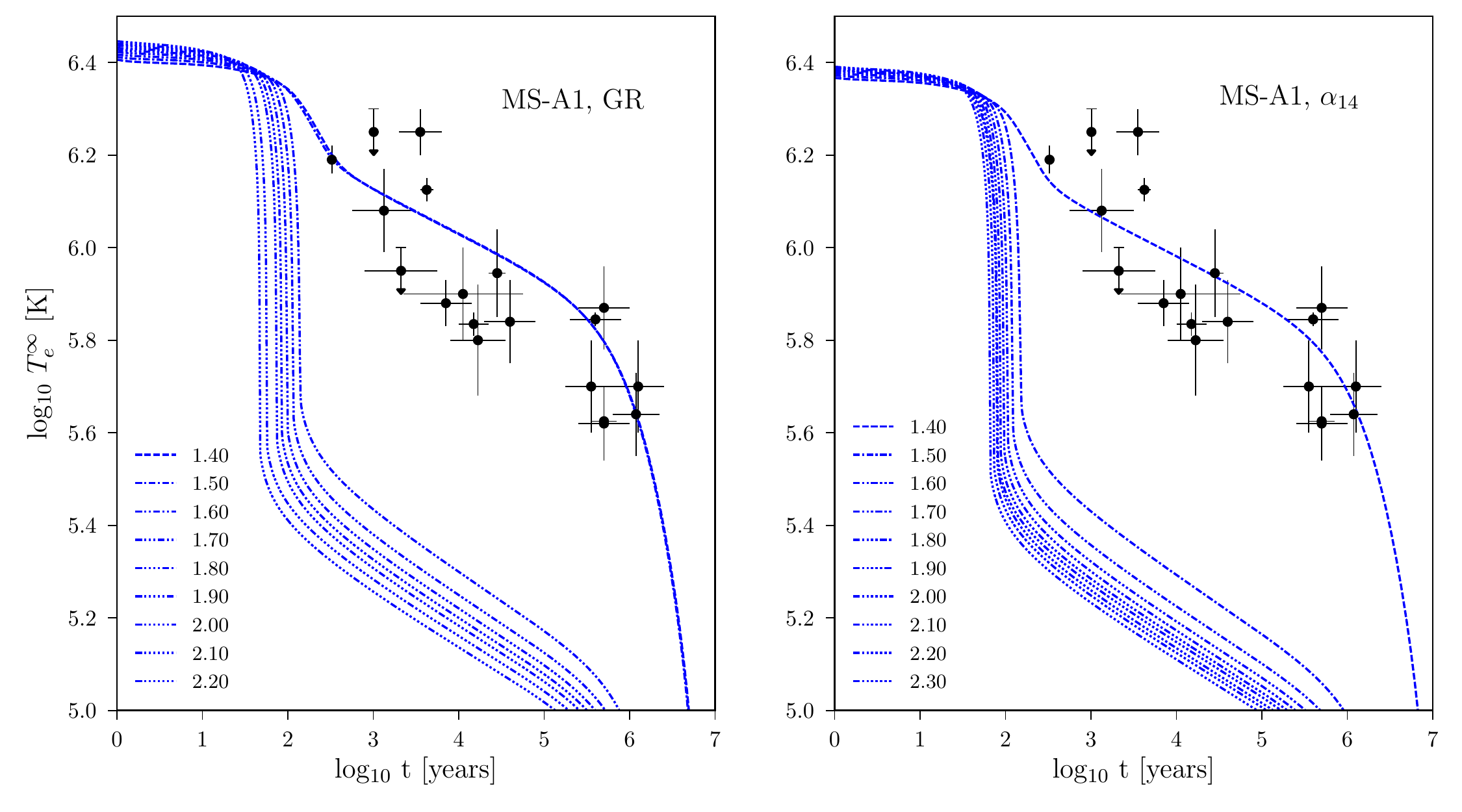}
	\caption{Cooling curves for the MS-A1 EoS, considering (Left panel) GR; (Right panel) $\alpha_{14}R^2$ gravity.
	 Superfluidity is absent in both panels, and $\eta$ is fixed to $10^{-20}$.}
	 \label{fig:cool2} 
\end{figure*}

As a first step in our exploration we consider a 1.4 $M_\odot$ neutron star model built with the APR EoS, without superfluidity/superconductivity
effects and employing an envelope containing only heavy elements.
The cooling of such a star is shown in the left panel of Fig.~\ref{fig:cool1} comparing the GR case with five modified gravity cases.
We see a very rough agreement between the predicted $T_e^\infty$ and the observational data.
Better agreement with the data can be reached with more sophisticated models that will be presented below.

In standard GR cooling, one distinguishes two different eras, named according to the particle species that dominate energy losses.
Initially we have the \textit{neutrino cooling era} lasting till an age of $\sim 10^{5-6}$ yrs at which time the slope of the $T_e^\infty - t$ curve becomes steeper, indicating the beginning of the \textit{photon cooling era}.
These epochs are clearly visible in the $\alpha R^{2}$ models as well. 
During the first era we see that increasing $\alpha$ results in models with lower $T^{\infty}_{e}$ while 
during the second era this behavior slightly inverts, yielding less steep slopes.
Notice that during the first $~100$ years, the ``plateau'', very little evolution is seen as this phase corresponds to the thermal relaxation of the crust, where 
neutrino emission is weak, resulting in a stationary $T_e^\infty$.

To understand the difference in cooling behavior appearing in the left panel of Fig.~\ref{fig:cool1} when gravity is modified,
we notice that it is unlikely due to a change in the deepest part of the star 
since for APR and a 1.4 $M_\odot$ star the central density practically does not change with gravity, as seen in the right panel of Fig.~\ref{fig:mreosa}.
However, the left panel of the same figure shows that the radius of the star significantly increases with $\alpha$ which itself implies a decrease of the
surface gravity, exhibited in the right panel of Fig.~\ref{fig:gserr}.
This simple change of $g_s$ has an immediate effect on $T_e$ through the thickening of the envelope (see below Eq.~\ref{eq:Lbund}).
In the right panel of Fig.~\ref{fig:mreosa} we present the same models but where the value of $g_s$, as used in the envelope models, has been artificially fixed to its GR value, demonstrating that change in $g_s$ explains most of the effect of modified gravity in these simple models.

Having explained the similarities and differences that both gravity theories exhibit in simple cooling curves, 
in Fig.~\ref{fig:cool2} we compare a larger set of NS mass-models for the MS-A1 EoS. 
Among the covered masses, in GR the 1.4 and 1.5 $M_\odot$ models are below the critical $M_\mathrm{DU}$ and the dramatic effect of the DUrca process
is clearly seen in all more massive stars.
During the early plateau $T_e^\infty$ is controlled by the crust and is not yet sensitive to the cooling of the core and the effect of the DUrca process only
appears at the end of this phase.
In the case of the $\alpha_{14}$ models, their behaviors are very similar to the GR case, with the significant difference that 
the critical mass $M_\mathrm{DU}$ is lower than in GR resulting in the 1.5 $M_\odot$ model already exhibiting fast DUrca cooling.

The dependence of  $T_e^\infty$ on $g_s$ explains the temperature spread seen during the plateau as $g_s$ increases when $M$ increases from 1.4 to
2.2 $M_\odot$ (in GR) or 2.3 $M_\odot$ (in $\alpha_{14}$).
However, this increase in $g_s$ is smaller in the $\alpha_{14}$ case than in GR, see right panel of Fig.~\ref{fig:gserr}, resulting in a smaller spread in 
plateau effective temperatures.
Moreover, during the plateau, the $\alpha_{14}$ models are globally slightly less warm than the GR models 
because their $g_s$ are globally slightly smaller.

After the plateau, this $g_s$ effect is still active and visible in the low mass models, below the DUrca threshold, which are cooler in the $\alpha_{14}$ case
than in the GR case.
However, in the models with the DUrca process acting this effect is counteracted by the fact that the $\alpha_{14}$ models have
the DUrca process acting in a smaller volume (see right panel of Fig.~\ref{fig:duveos}) and hence have a smaller neutrino luminosity resulting in
a higher temperature.

\subsection{Realistic cooling models: superfluidity/superconductivity and light element envelopes}

\begin{figure*}[t]
	\includegraphics[scale=0.65]{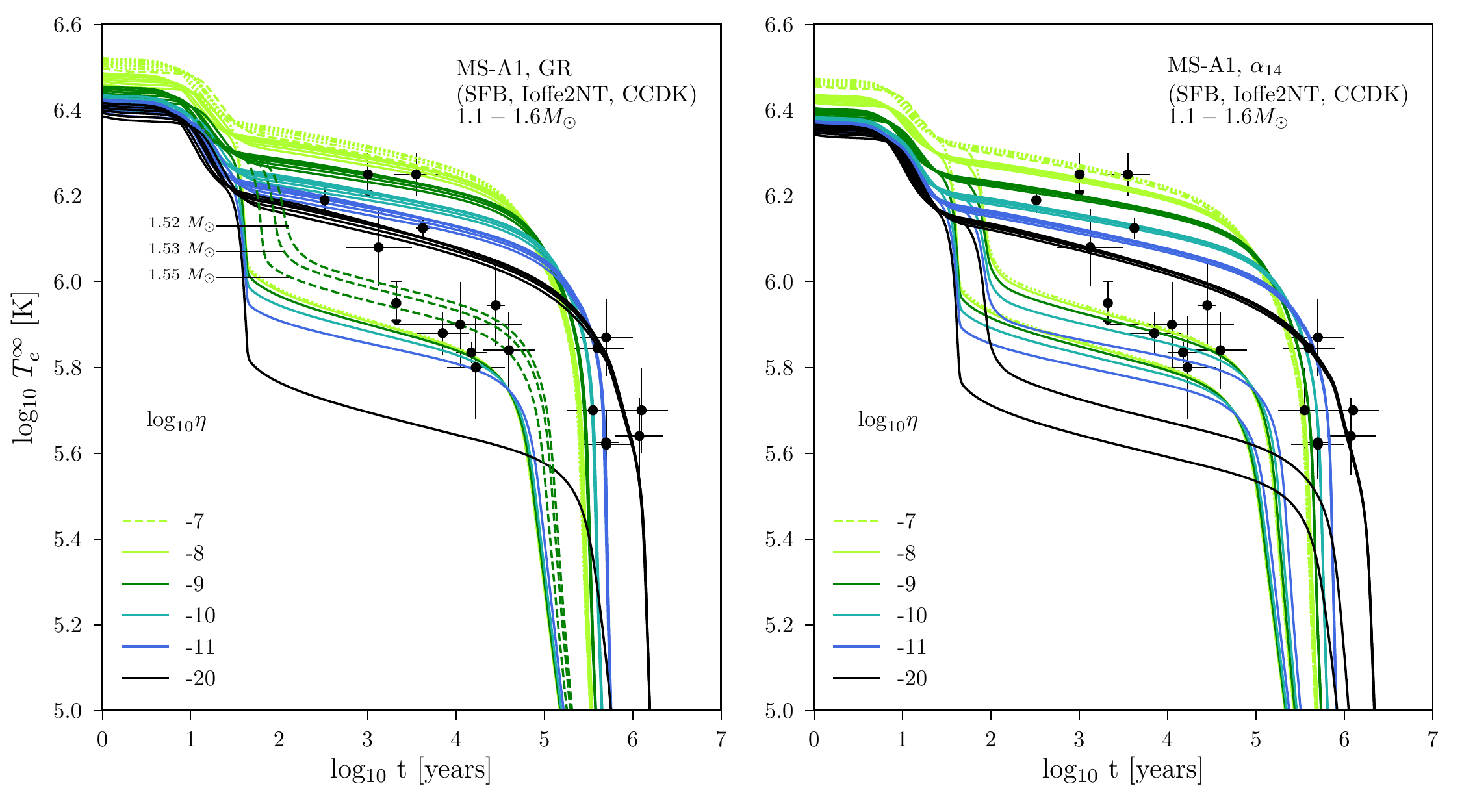}
	\caption{Cooling curves for the MS-A1 EoS, considering variable $\eta$ values and the indicated superfluidity gaps. 
	Left panel: GR. Right panel: $\alpha_{14}R^2$.}
	\label{fig:cool3} 
\end{figure*}

After the presentation of very simple models in the previous section we are now in position to consider a first case study of realistic scenarios in which two additional
essential ingredients are present, namely pairing which can strongly suppress neutrino emission and the presence of light elements in the envelope
which significantly alters the observable $T_e^\infty$.
We will compare models in GR and in $\alpha_{14}$ built with the MS-A1 EoS, considering masses between 1.1 and 1.6 $M_\odot$ in steps of $0.1 M_\odot$.
Notice that in the GR case models up to $1.5 M_\odot$ have slow neutrino emission and only the $1.6 M_\odot$ model has the DUrca process allowed, 
while in the $\alpha_{14}$ both $1.5$ and $1.6 M_\odot$ models have the DUrca process acting.
Models with higher mass result in lower temperatures than the $1.6 M_\odot$ one and we do not include them to avoid saturating the figures.

We present in Fig~\ref{fig:cool3} the cooling of these more complex models, and a first looks shows that the predicted range of $T^{\infty}_{e}$ is broader than those of the previous section (see Fig.~\ref{fig:cool2}), effectively covering the inferred values for the 19 NSs of our adopted data set.

The vertical spread of the cooling curves with models of mass below $M_{\mathrm{DU}}$, which arises from the varying amount of light elements in their envelopes, allows to fit the hottest stars on the sample. 
On the other hand, models with $M\geq M_{\mathrm{DU}}$ in which the fast neutrino emission of the DUrca process has been controlled by pairing
can explain the colder objects.

\begin{figure*}[t]
	\includegraphics[scale=0.6]{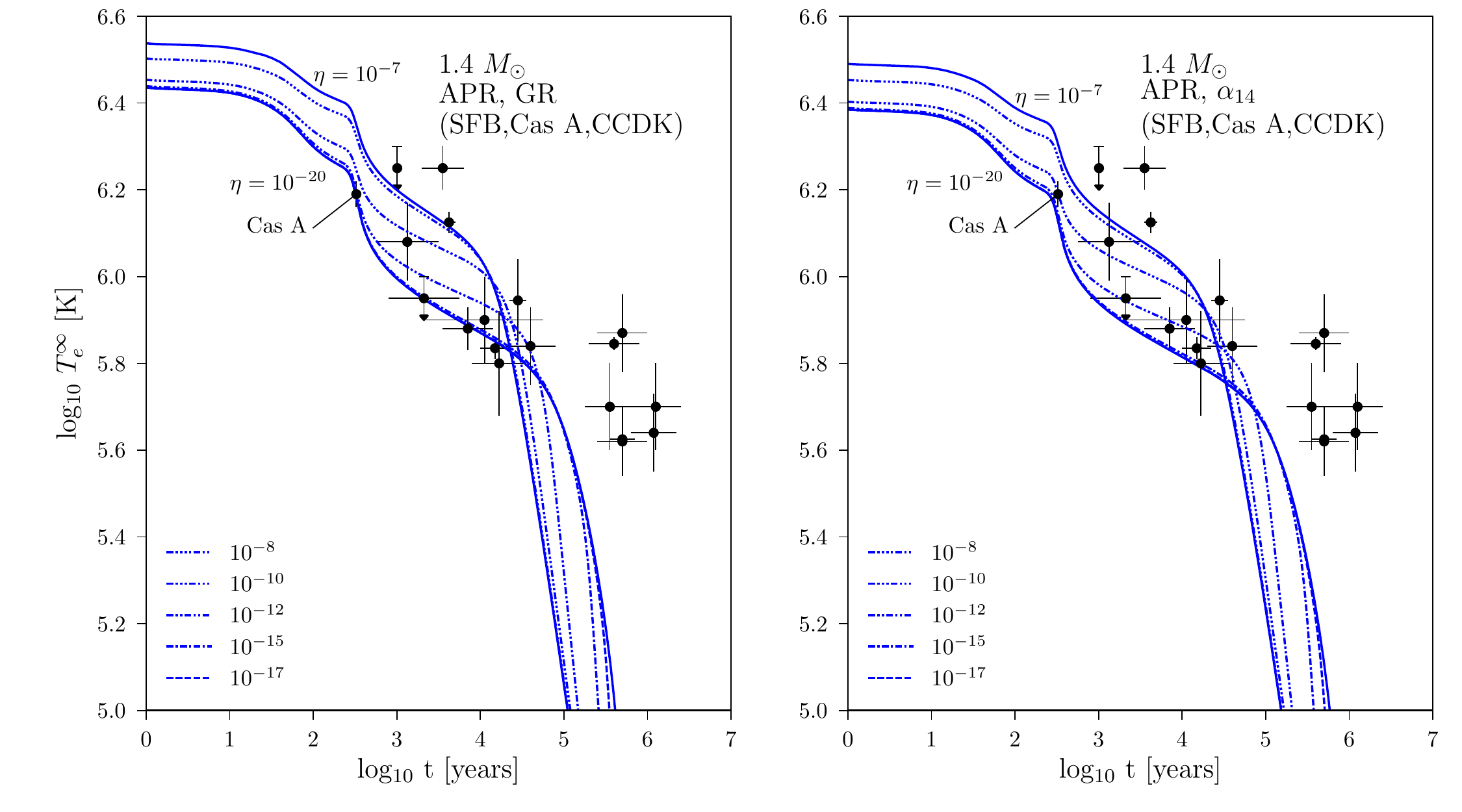}
	\caption{Comparison of a 1.4$M_{\odot}$ APR-NS model with several values of $\eta$ in either GR (left panel) and $\alpha_{14}R^2$ gravity.
	The superfluidity/superconductivity gaps employed are indicated on each panel and were chosen to induce a rapid cooling of the star at the age
	of the Cas A neutron star.}
	\label{fig:cool4} 
\end{figure*}

Let us consider in more detail the effects of increasing the amount of light elements in the envelope, parametrized by $\eta$ in Eq.~\ref{eq:eta}, by
considering the neutrino cooling era of stars with masses below $M_\mathrm{DU}$.
We have five bands of cooling curves for the adopted five values of $\eta$, each band encompassing masses from 1.1 to 1.5 $M_\odot$ in GR and
1.1 to 1.4 $M_\odot$ in $\alpha_{14}$: varying the mass has little effect, more massive stars being only slightly cooler than less massive ones, 
while increasing $\eta$ significantly raises $T_e^\infty$ because of the higher thermal conductivity of light elements compared to iron-peak ones.
Within such a type of scenario the spread of $T_e^\infty$ in hot young NSs is a result of their having accreted different amounts of matter in their past,
either when they were formed in a supernova explosion or during their later evolution.
Notice that during the neutrino cooling era models with different values of $\eta$ have identical internal temperatures, since neutrino losses control 
this temperature, and the ones with larger $\eta$ just appears warmer at the surface but when entering the photon cooling era the situation in reversed:
models with larger $\eta$ will emit more photons that result in a faster cooling.
This inversion is clearly seen in both in the GR and the $\alpha_{14}$ cases and one can also notice that models with larger $\eta$ and larger photon luminosity
naturally enter the photon cooling era earlier that models with lower $\eta$.

The spectacular effect of suppression of the fast DUrca neutrino emission is clearly seen by comparing Fig~\ref{fig:cool2} with Fig~\ref{fig:cool3}:
in the first case, all models with the DUrca process acting have $T_e^\infty$ well below all observed values while once the neutrino emission is
controlled by pairing, $T_e^\infty$ comparable to observed ones are easily obtained. 
Together with the effects of the presence of light elements in the envelope, we are able to generate families of models that easily cover the
range of $T_e^\infty$ for the coldest observed young neutron stars.

We emphasize that the neutron star mass range that describes these cold young stars, with ages below $10^5$ years, depends on the assumed EoS.
In the case of the APR EoS, with $M_\mathrm{DU} = 1.97 M_\odot$, models with masses above $2 M_\odot$ are needed to produce cold young stars.

Finally, comparing the two panels of Fig~\ref{fig:cool3}, it is clear that the small differences in cooling behaviors when altering the gravity theory from GR
to $\alpha R^2$ are much smaller that the effects of either microphysical (as, e.g., pairing) or astrophysical (as mass or envelope composition) ingredients.

\subsection{Cooling of the Cassiopeia A neutron star}
\label{sec:CasA}

The neutron star in the Cassiopeia A supernova remnant was discovered in the Chandra first light observation \cite{Tananbaum:1999aa}.
It is the youngest known neutron star, its supernova remnant having a kinematic age of $\sim 340$ yrs \cite{Fesen:2006tx} that is supported by
a possible historical observation of the supernova by J. Flamsteed on August 16th, 1684 \cite{Ashworth:1980wd}.
Its soft X-ray thermal spectrum is well described by a non magnetized carbon atmosphere model \cite{Ho:2009wp} implying a surface temperature 
$\sim 2\times 10^6$ K.
Analysis of 10 years of Chandra observations of the supernova remnant pointed to a rapid cooling of the neutron star \cite{Heinke:2010vr},
its effective temperature having apparently decreased by $\sim 4\%$ during that time span.
Analyses of another decade of observations reached a similar conclusion \cite{Wijngaarden:2019vr,Shternin:2021vn}
while a different approach \cite{Posselt:2013ux,Posselt:2018tb} found a much smaller decrease in $T_e^\infty$, if any at all.

We here accept the possibility of rapid cooling of this neutron star and the interpretation of this evolution being due to a sudden increase in the
neutrino luminosity, through the PBF process, resulting from the initiation of the neutron $^{3}\! P_{2}$ superfluidity phase transition 
\cite{Shternin:2011ul,Page:2011wu}.
In this scenario, protons become superconducting at an early time, suppressing neutrino emission from all processes involving protons
 (modified Urca, n-p and p-p bremsstrahlung, see Table~\ref{Tab:nu}), thus slowing the early cooling of the star. 
 Due to this slow cooling, the core of the neutron star reached the neutron $^{3}\! P_{2}$ superfluidity critical temperature only very recently, 
 implying a $T_c \sim 5\times 10^8$ K \cite{Shternin:2011ul,Page:2011wu,Shternin:2021vn}.

We display in Fig.~\ref{fig:cool4} this model in GR and in $\alpha_{14}$ modified gravity considering a star with a total gravitational mass of $1.4 M_\odot$
and pairing gaps chosen to produce the desired cooling behavior.
As seen previously,  $\alpha_{14}$ models are slightly colder than GR ones but, by considering variants with different amounts of light elements in the envelope,
the same scenario can result in a rapid cooling of the neutron star at its present age and with a $T_e^\infty \simeq 1.55 \times 10^6$ K, in both cases:
$\eta \lesssim 10^{-12}$ in GR and $\eta \sim 10^{-10}$ in $\alpha_{14}$.

\subsection{Cooling of NS 1987A}
\label{sec:NS1987A}

\begin{figure*}[t]
	\includegraphics[scale=0.7]{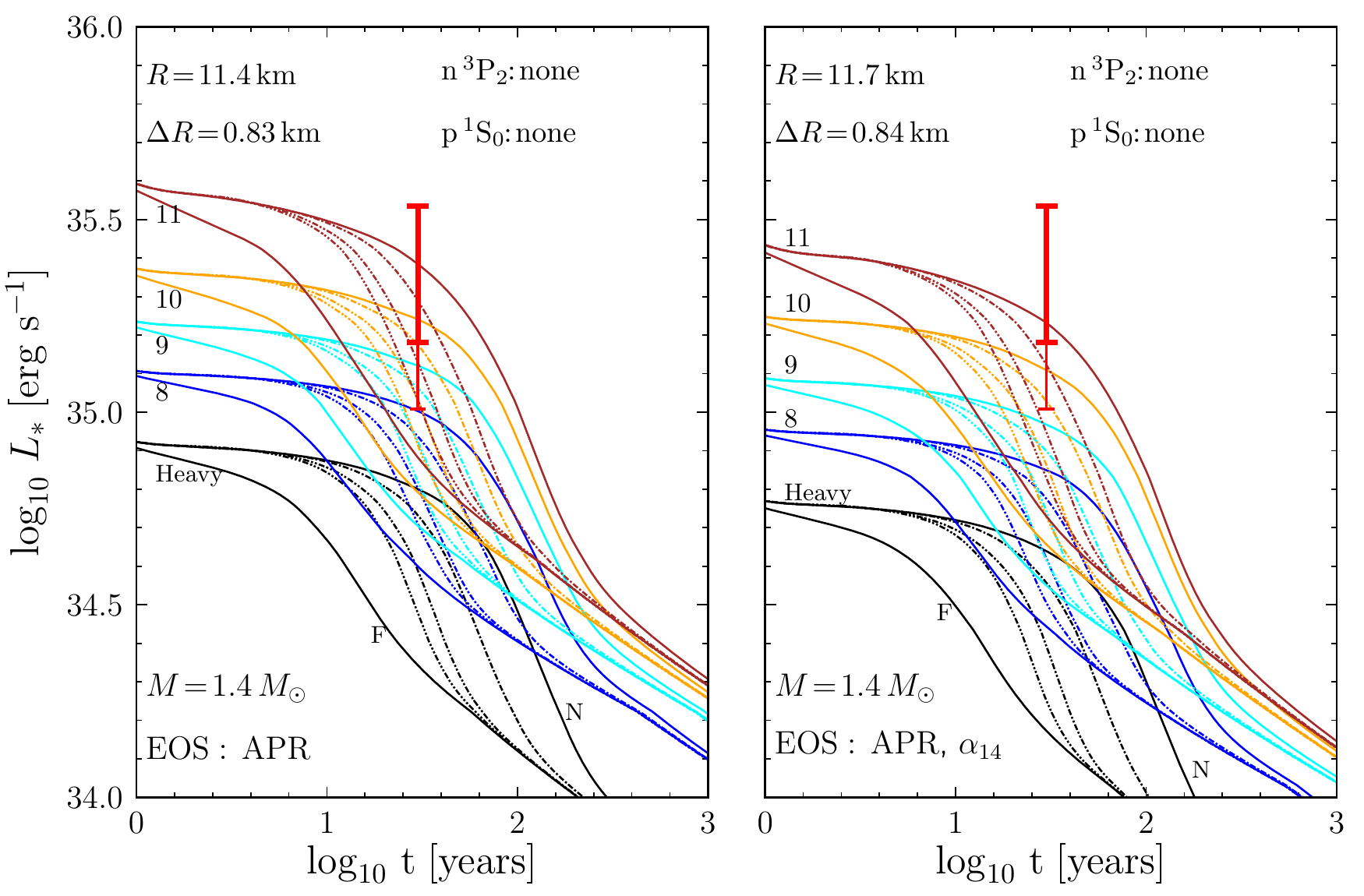}
	\caption{Cooling curves for a 1.4 $M_{\odot}$ APR NS model. Left panel: GR. Right panel: $\alpha_{14}R^2$ gravity. 
	In each panel the (red) vertical bar marks the $1\sigma$ range of the inferred surface thermal luminosity of the putative neutron star NS 1987A
	(the lower extension of the bar extends the range when taking into account pre-heating of the dust by $^{44}$Ti decay).
	In each case five families of cooling models are presented that have different amount of light elements in their envelopes, 
	ranging from no light elements and marked as ``Heavy'' and light elements present up to a density of 
	$\rho_\mathrm{L} = 10^\mathrm{L}$ g cm$^{-3}$ with $L=8, 9, 10,$ and 11, as labelled.
	Within each family, the line-style corresponds to a different assumed neutron $^{1}\! S_{0}$ superfluid gap:
	continuous lines correspond to assuming inner crust neutrons are either fully superfluid, labelled as ``F'' in the heavy element envelope case, 
	or not superfluid at all, labelled as ``N'', while the three other lines have line-style as in Fig.\ref{fig:Tc}.}
	\label{fig:cool5} 
\end{figure*}

As a last case study we consider the possible recent discovery of the neutron star NS 1987A produced by the supernova SN 1987A,
whose putative existence was revealed by the presence of a blob of warm dust close to center of explosion \cite{Cigan:2019ur,Page:2020wj}. 
From the luminosity of the blob a luminosity $\sim 1-3 \times 10^{35}$ erg s$^{-1}$ is inferred for the neutron star.
With an age of 28 yrs at the time of the discovery observation, this very young object allows us to probe the thermal evolution of a neutron star
during the early plateau.
This stage of evolution is quite insensitive to the physics of the core and the dominant characteristics that influence it are: in the first place, the thickness of the crust
which geometrically determines the cooling time scale; secondly, the extent of $^{1}\! S_{0}$ neutron superfluidity in the inner crust
which can strongly reduce the specific heat and fasten the cooling, and, finally, the thickness of the light elements layer in the envelope
which controls the emerging photon luminosity.
Within GR, this scenario was studied in detail in \cite{Page:2020wj} and we reproduce in Fig.~\ref{fig:cool5} one set of models from that work and compare them 
with their $\alpha_{14}R^2$ gravity counterparts.
The crust thickness is little affected by the change in the gravity models for stars of mass $\sim 1.4 M_\odot$ as we showed in Fig.~\ref{fig:crustt}
and the only gravity effect of relevance is the reduction of the surface mass $M_\mathrm{surf}$ which controls the thickness of the envelope and the light
element layer: 
comparing the two panels of Fig.~\ref{fig:cool5} one sees that the only significant difference is an almost uniform reduction of the luminosity
in the $\alpha_{14}R^2$ models compared to their GR counterparts, 
the same effect we have identified as a reduction of $T_e^\infty$ in \S~\ref{sec:cooling_simple}.
As a result, the interpretation of the warm dust blob as being a sign of the presence of NS 1987A is still sustainable within the $\alpha R^2$ theory, 
but an extreme modification as in the $\alpha_{14}$ case requires a significantly thicker light element surface layer:
in GR gravity a light element layer reaching a density of $\rho_\mathrm{L}  \sim 10^9$ g cm$^{-3}$ results is luminosities comparable to the lower requirement of $L_* \sim 10^{35}$ erg s$^{-1}$
while within the extreme case of  $\alpha_{14} R^2$ gravity $\rho_\mathrm{L}  \sim 10^{10}$ g cm$^{-3}$ is implied.
However, the analysis of Ref. \cite{Page:2020wj} showed that, even if light elements as O may avoid destruction by nuclear burning to such high densities as $\rho_\mathrm{L}  \sim 10^{10}$ g cm$^{-3}$
in the hot environment of a young neutron star envelope (see Fig. 3 of that paper),
how such a thick layer of light elements may have been deposited is unclear since even accretion at the Eddington rate (i.e., at about $10^5$ g cm$^{-2}$ s$^{-1}$) 
would have taken about 20 yrs and the resulting luminosity, $L_\mathrm{Edd} \sim 10^{38}$ erg s$^{-1}$, would have exceeded observational upper limits on the luminosity of that object \cite{Alp:2018ao}.

\section{Discussion}
\label{sec:discussion}

In this work we have studied in detail how a shift in the description of gravity from General Relativity to an $f(R)$ type of theory,
in the particular case of $f(R)=R+\alpha R^2$, modify our description of the structure and thermal evolution of neutron stars.
Conversely, we also examined the potentialities for constraining such modification of gravity theory using neutron star observables.
With regard to the structure of dense matter above $10^{13}$ g cm$^{-3}$ \cite{Burgio:2021vj}, 
due to the uncertainties on the nature of strong interactions at high densities
and the limitations of many-body theory in this regime, we employed four different EoSs that cover the present range of expectations 
in the case this matter consists only of nucleons and leptons \cite{Han:2019aa,Page:2020wj}.
Specifically, the EoS MS-A1, MS-B1, and MS-C1, were designed so that, together with the APR EoS, 
the deduced radius of a $1.4 M_\odot$ neutron star is between 11.6 and 13.2 km,
as the latest observational data indicate is the most likely range \cite{Raaijmakers:2021vc}, 
and all four have a maximum mass above $2.1 M_\odot$ in agreement with the most massive pulsar mass measurements, 
$1.97 \pm 0.04 \, M_\odot$ for PSR J1614-2230 \cite{Demorest:2010ts}, 
$2.01 \pm 0.04 \, M_\odot$ for PSR J0348+0432 \cite{Antoniadis:2013wk}, and
$2.17 \pm 0.11 \, M_\odot$ for PSR J0740+6620 \cite{Cromartie:2020vg}. We do not expect the consideration of EoSs with more degrees of freedom, as, e.g., the inclusion of hyperons or deconfined quarks, to
have any significant impact on our conclusions as we discuss below.

As a first step we studied global properties of neutron stars, their mass and radius.
Three different mass concepts appear, the baryonic mass, $M_\mathrm{bar}$, and the total and surface, $M_\mathrm{tot}$ and $M_\mathrm{surf}$, gravitational masses.
The latter two are obtained from the metric, Eq.~\ref{eq:mr}, and are identical in GR while $M_\mathrm{surf} < M_\mathrm{tot}$ in $\alpha R^2$ theories.
Neutron star mass measurements are typically obtained from binary orbital motion, i.e. in the regime $r \gg r_*$, and are thus providing us with $M_\mathrm{tot}$.
For this reason we label all our neutron star models with the total gravitational mass of the model.
The baryonic mass is not directly measurable but is an intrinsic property of the neutron star that is independent of gravity, being just its baryon number multiplied by a baryonic mass.
The surface gravitational mass is potentially measurable as it should strongly affect the bending of photon trajectories nearby the stellar surface 
 \cite{Pechenick:1983ts,Page:1995vi,Leahy:1995vl} and
light curve modeling may be able to find differences between $M_\mathrm{surf}$ and the
binary motion inferred $M_\mathrm{tot}$ in the case of very accurate modeling as possible with NICER data (see, e.g., Ref. \cite{Bogdanov:2021wh}).
We found, see Fig.~\ref{fig:mbarionic},  that 
{\it the difference between $M_\mathrm{tot}$ and $M_\mathrm{surf}$ depends only on $\alpha$ and almost nothing else} for non-rotating stars,
at least within the range of measured gravitational masses and the presently restricted range of EoSs.
This difference can reach half a Solar mass in the extreme case of  $\alpha_{14} R^2$ gravity opening a promising road to confirm of refute such modifications of gravity.

With respect to the effects of modified gravity on the mass-radius and mass central-density relationships our results, 
presented in the Figures \ref{fig:mreosa} and \ref{fig:mreosb}, confirmed results of previous authors, e.g., \cite{Yazadjiev_2014, Feola:2020tw}.
The main effect is that, independently of the employed EoS, larger gravitational masses are reachable within $\alpha R^2$ gravity theories.
Stellar radii are larger, except for very low mass stars, and central densities are lower, for high mass stars, in $\alpha R^2$ gravity stellar models
than in the corresponding GR models with the same $M_\mathrm{tot}$ but what is precisely meant by ``very low mass'' and ``high mass'' is dependent on the EoS.

We considered finer details of the internal structure, as the total physical volume of the star, Fig.~\ref{fig:gsduvapr}, the thickness of the crust,
Fig.~\ref{fig:crustt}, the surface gravitational acceleration $g_s$, and the threshold mass for the direct Urca neutrino emission process.
In the cases of the first two quantities we found no difference between GR and $\alpha R^2$ gravity that we could expect to lead to some clear observable signal.
A reduction of $g_s$ by up to a factor of two is possible, see Fig.~\ref{fig:gserr}, and its consequences are discussed below in the context of thermal evolution.
About the direct Urca process, its occurrence is determined by the matter density being above some EoS dependent critical density $\rho_\mathrm{DU}$
and any star whose central density $\rho_0$ is above $\rho_\mathrm{DU}$ can undergo very fast neutrino cooling.
A possible observable, or observationally deducible, quantity would be the critical total gravitational mass $M_\mathrm{DU}$ at which
 $\rho_0$ reaches $\rho_\mathrm{DU}$.
In the case $\rho_\mathrm{DU}$ is around or below $10^{15}$ g cm$^{-3}$, the difference in $M_\mathrm{tot}$ between GR and $\alpha R^2$ gravity
is quite small and thus $M_\mathrm{DU}$ is not significantly altered by modifying gravity.
In contradistinction, in the case of an EoS for which $\rho_\mathrm{DU}$ is well above $10^{15}$ g cm$^{-3}$, as in our two chosen EoSs  APR and MS-C1, 
$M_\mathrm{DU}$ is increased by up to 0.2 $M_\odot$ in $\alpha_{14} R^2$ gravity compared to GR:
at present time the actual value of $\rho_\mathrm{DU}$ is uncertain, as exemplified by the range it covers in our four chosen EoS, 
but, being optimistic, future progress in both dense matter theory to constrain $\rho_\mathrm{DU}$ and 
mass measurements of some neutron stars undergoing fast neutrino cooling
could provide us with a handle to constraint modifications to gravity theory in the case $\rho_\mathrm{DU}$ turns out to be large.
Fast neutrino emission through some form of direct Urca process can also be the result of the presence of deconfined quark matter, hyperons, 
or charged meson condensates (see, e.g., \cite{Page:2006tr} for a brief description) and in all these cases there is a also corresponding critical density
so that the same considerations apply.
Evidence for fast neutrino cooling can be found from neutron stars in transiently accreting binary systems \cite{Colpi:2001vw,Beznogov:2015wn}
for which mass measurements, or at least mass estimates, are possible.
At present time a few such systems have been identified, as 
SAX J1808-3658 \cite{Wijnands:2002wf,Campana:2002uj}, 1H 1905+000 \cite{Jonker:2006wl}, MXB 1659-29 \cite{Brown:2018tp}, 
and HETE J1900.1-2455 \cite{Degenaar:2021us}, but none of them has mass measurement.

Turning now to the impact of modified gravity on the cooling of neutron stars we first showed in \S~\ref{sect:cooleqs} that the thermal evolution equations
retain the same from in $\alpha R^2$ gravity as they have in GR.
This has the convenient implication that a neutron star cooling code such as \texttt{NSCool} can be immediately used without any change.
In a first, very simple scenario, we exhibited in Fig.~\ref{fig:cool1} a trivial effect of $\alpha R^2$ gravity in that the reduction in $g_s$ 
results in slightly lower surface temperatures during the neutrino cooling era.
However, even small changes in the envelope chemical composition, as displayed, e.g., in Fig.~\ref{fig:cool3}, can induce a rise in the surface temperature
during the neutrino cooling era that can be much more important that the small decrease due to the $\alpha R^2$ gravity reduction in $g_s$.
This effect is thus completely hidden by the ``noise'' from the uncertainty in the envelope chemical composition for which we usually have almost 
no constraint. 
Considering a set of stellar models covering a range of masses we showed Fig.~\ref{fig:cool2} the immediate effect that a model, the 1.5 $M_\odot$ one 
in this particular example, that undergoes slow neutrino cooling in GR gravity because its mass is below $M_\mathrm{DU} = 1.509 M_\odot$,
will experience fast neutrino cooling in $\alpha_{14} R^2$ gravity because its mass is now above new value of $M_\mathrm{DU}$ 
which as decreased to 1.474 $M_\odot$.
As long as no mass measurements are available for isolated neutron stars such an effect is unfortunately only of academic interest and, moreover,
a small change in $\rho_\mathrm{DU}$ due to a small change in the EoS properties can have the same effect of modifying $M_\mathrm{DU}$.

After these first results, we  considered three case studies for a better evaluation of the magnitude of the impact of modifying the gravity theory
on our ability to infer properties of dense matter from such studies.
In the first case we revisited the scenario \cite{Page:1990vc,Page:1992tz} in which fast neutrino emission, which results in very cold neutron stars if uninhibited,
is controlled by pairing. 
This scenario is well known (see, e.g., the reviews \cite{Yakovlev:2004iq,Page:2006vt} and the recent works \cite{Beznogov:2015wn} and \cite{Beznogov:2015vm})
to result in a wide range of predicted effective temperatures that allow to interpret all present data on isolated cooling neutron stars
and we showed in Fig.~\ref{fig:cool3} that the same conclusion holds even within the most extreme case of $\alpha_{14} R^2$ gravity theory.
Since increasing $\alpha$ has no further noticeable effect compared to $\alpha_{14}$ while at smaller $\alpha$ the modifications are even smaller,
we can conclude that the validity of this family of cooling scenarios, fast neutrino emission controlled by pairing, is not altered by modification of the
gravity theory within the $\alpha R^2$ scheme.

In the second case we revisited the intriguing possible observation of the cooling of an isolated neutron star in real time, the neutron star in the 
supernova remnant Cassiopeia A.
We considered the specific scenario of sudden increase of the star's neutrino luminosity due to the very recent onset of neutron $^{3}\! P_{2}$ pairing in its core
\cite{Shternin:2011ul,Page:2011wu} and found, as shown in Fig.~\ref{fig:cool4},
that a small increase in the assumed thickness of the light element 
layer in its envelope, an essentially unknown quantity, is sufficient to counterbalance the effect of the change in gravity theory
as already mentioned in the above described simple models.
As a result, the underlying physical interpretation of the observation as due to the appearance of superfluidity is not affected at all. 

Our last case study is the analysis of the recent identification of a likely first sign of the neutron star, dubbed NS 1987A,
created in the SN 1987A supernova event \cite{Cigan:2019ur,Page:2020wj}.
A key point of the argumentation is that the expected thermal luminosity of a $\sim 30$ years old neutron star is sufficient to energize the dust blob
suspected to conceal NS 1987A which needs an energy input of at least $10^{35}$ erg s$^{-1}$.
Producing such a high thermal luminosity was shown to be possible, within GR gravity, in the case the outer layer of the neutron star contain a significant amount of light elements, as C or O \cite{Page:2020wj}.
In the case of $\alpha R^2$ gravity, as shown in \S~\ref{sec:cooling_simple}, slightly lower thermal luminosities are obtained with the consequences for NS 1987A, as shown in Fig.~\ref{fig:cool5},
that a thicker layer of light elements is needed compared to the requirements of analogous models in GR gravity.
In the case of $\alpha_{14} R^2$ gravity, light elements would need to be present up to a density $\rho_\mathrm{L}  \sim 10^{10}$ g cm$^{-3}$ that may be difficult to reach as we discussed in 
\S~\ref{sec:NS1987A}.
If the existence of NS 1987A, and with such a high thermal luminosity, is confirmed, it may offer us an interesting object to test gravity theory.

When finishing the present work we became aware of a similar work by Dohi et al. \cite{Dohi:2021uq}  in the context of the very similar theory 
$f(R) = R + R^{4/3}$.
The difference between both theories appears to be small and the differences in the results are very small.
In the case were cooling curves in GR and $\alpha R^{4/3}$ gravity are compared, as in their Fig. 11, 13, and 15, their results are in perfect agreement with ours.
These authors label their neutron star models using $M_\mathrm{surf}$ instead of $M_\mathrm{tot}$:
as a result they find that the threshold mass $M_\mathrm{DU}$ can be much smaller in $\alpha R^{4/3}$ gravity than in GR.
However, once employing the total mass $M_\mathrm{tot}$ as we do, considering this to be the measured one when a measurement is available,
the difference is actually much smaller.
In their conclusion, these authors mention that the crust relaxation time, which they refer to as the ``knee'', can be affected by gravity theory and
be used as a probe for it, a claim we strongly disagree with: 
we found that the crust thickness, a major ingredient in the crust relaxation time, is little affected by gravity changes, at least with the scheme we studied here.
Moreover, there are many physical ingredients in the crust that are still poorly determined and have a much larger effect
than changing the underlying theory of gravity, as we have shown in our Figs.~\ref{fig:cool1}-\ref{fig:cool5} and has been described with much detail within GR in, e.g. \cite{Page:2009um}.

\section{Conclusions}
\label{sec:conclusions}

We have presented an extensive study of the structure and evolution of isolated neutron stars within the $f(R) = R+\alpha R^2$ extension of gravity theory 
beyond the GR case of $f(R)=R$.
We found, as was already hinted by previous authors who only studied the structure, that the deviations from GR saturate in both the structure and the evolution 
when $\alpha$ grows and theories with $\alpha\geq 10^{14}$ cm$^{-2}$ result in essentially identical predictions.
This result has the important consequence that the whole range of predictions from this family of gravity theories can thus be covered 
by actually scanning only a finite range of values of $\alpha$.

For a given high density EoS, predicted values for the maximum mass and the radii of stars with masses above about $1 M_\odot$ are larger within $\alpha R^2$
gravity theories than within GR gravity.
However, the changes when going from GR to $\alpha R^2$ gravity are comparable in size to changes that are obtained within GR simply by considering different EoSs.
Thus, using neutron star mass and radius measurements to constrain theories of gravity do not seem to be a promising aproach,
as long as present uncertainties on the high density EoS remain.
Nevertheless we have been able to identify one property, the surface gravitational mass $M_\mathrm{surf}$, which is potentially measurable through light curve modeling,
and differs in $\alpha R^2$ gravity from the total gravitational mass $M_\mathrm{tot}$ that is measured from binary orbit,
while both masses are identical in GR gravity.
Most importantly, we have shown, see Fig. \ref{fig:mbarionic}, that the difference $M_\mathrm{tot}-M_\mathrm{surf}$ depends only on $\alpha$ and 
turns out to be almost independent of  $M_\mathrm{tot}$ and of the high density EoS.
{\it Measuring this difference would be a direct measurement of $\alpha$, within the $\alpha R^2$ gravity scheme}.

Regarding the thermal evolution of neutron stars, many physical and astrophysical ingredients beyond just the EoS are needed for a realistic description.
We have shown that uncertainties in these many ingredients have stronger impact in the models than the change in structure resulting from a change in gravity
theory, at least with the $\alpha R^2$ gravity scheme.
By studying a few well-known cooling scenarios we have shown that modified gravity within the  $\alpha R^2$ scheme alters neither their validity nor the conclusions
that may be derived from them regarding properties of dense matter.
We nevertheless found one case, the neutron star NS 1987A in the remnant of SN 1987A, whose deduced high thermal luminosity requires the presence of
a very thick layer of light elements at its surface: within $\alpha R^2$ gravity with large $\alpha$ the presence of an unrealistically thick layer of light elements may be
required resulting in a tension between such gravity theories and the interpretation of the ALMA detected warm blob of dust luminosity being due to the
heating from the young neutron star \cite{Page:2020wj}.

\begin{acknowledgments}
This work was supported by the Mexican Consejo Nacional de Ciencia y Tecnolog{\'\i}a with a CB-2014-1 grant $\#$240512 and the 
Universidad Nacional Aut\'onoma de M\'exico through a UNAM-PAPIIT grant \#109520.
We thank S. Mendoza for his valuable commentaries and corrections on the original manuscript.
\end{acknowledgments}

\bibliography{Referencias}
\bibliographystyle{apsrev4-2}

\end{document}